\documentclass[12pt,preprint]{aastex}
\usepackage[dvips]{color,epsfig,rotating}
\usepackage{graphics}

\begin{document}

\title{The Luminosity Function of QSO Host Galaxies}

\author{Timothy S. Hamilton,\altaffilmark{1,2,3,4}
Stefano Casertano,\altaffilmark{2,4}
and David A. Turnshek\altaffilmark{1,4}}

\altaffiltext{1}{Dept. of Physics \& Astronomy, University of Pittsburgh,
Pittsburgh, PA 15260, USA}

\altaffiltext{2}{Space Telescope Science Institute, 3700 San Martin Drive, 
Baltimore, MD 21218, USA}

\altaffiltext{3}{Present address: Code 662, NASA/GSFC, Greenbelt, MD 20771, USA}

\altaffiltext{4}{email: hamilton@milkyway.gsfc.nasa.gov, stefano@stsci.edu, 
turnshek@pitt.edu}

\begin{abstract}

We present some results from our {\it HST} archival image study of 71
QSO host galaxies. The objects are selected to have $z\leq0.46$ and
total absolute magnitude $M_V\leq -23$ in our adopted cosmology
($H_0=50$ km s$^{-1}$ Mpc$^{-1}$, $q_0=0.5$, $\Lambda=0$). The aim of
this initial study is to investigate the composition of the sample
with respect to host morphology and radio loudness, as well as derive
the QSO host galaxy luminosity function. We have analyzed available
WFPC2 images in $R$ or $I$ band ($U$ in one case), using a uniform set
of procedures.  The host galaxies span a narrow range of luminosities
and are exceptionally bright, much more so than normal galaxies,
usually $L>L^*_V$.  The QSOs are almost equally divided among three
subclasses: radio-loud QSOs with elliptical hosts, radio-quiet QSOs
with elliptical hosts, and radio-quiet QSOs with spiral hosts.
Radio-loud QSOs with spiral hosts are extremely rare.  Using a
weighting procedure, we derive the combined luminosity function of QSO
host galaxies.  We find that the luminosity function of QSO hosts
differs in shape from that of normal galaxies but that they coincide
at the highest luminosities.  The ratio of the number of quasar hosts
to the number of normal galaxies at a luminosity $L_V$ is
$\mathcal{R}=(L_V/{11.48L^*_V})^{2.46}$, where $L^*_V$ corresponds to
$M^*_V=-22.35$, and a QSO is defined to be an object with total
nuclear plus host light $M_V\leq-23$. This ratio can be interpreted as
the probability that a galaxy with luminosity $L_V$ will host a QSO at
redshift $z\approx0.26$.

\end{abstract}

\section{INTRODUCTION}\label{sec:intro}

Much has been learned about the properties of QSO host galaxies since
they were first imaged almost three decades ago (Kristian~1973).
Early results include establishing a positive correlation between host
and nuclear QSO luminosities (Hutchings, Crampton, \& Campbell~1984)
and indications of a morphological difference between radio-loud and
radio-quiet QSOs, with the former more likely to be in elliptical
hosts and the latter in spiral hosts (Malkan, Margon, \& Chanan~1984).
Boroson, Persson, \& Oke~(1985), as well as Stockton \&
MacKenty~(1987) examine hosts to classify them spectroscopically and in
the context of their nuclear emissions.  Working in the near-infrared,
where the luminosity contrast is more favorable to the host galaxy,
Dunlop et al.~(1993) show that QSO hosts are typically drawn from the
bright end of the galaxy luminosity function (in agreement with
Hutchings et al.~1984).  McLeod \& Rieke~(1994a,b), also using
near-infrared data, find that hosts of radio-quiet QSOs are typically
represented by an exponential (spiral disk) light profile (in
agreement with Malkan et al.~1984), and that high-luminosity QSOs
generally have brighter hosts than low-luminosity QSOs (in agreement
with Hutchings et al.~1984).

High-resolution space-based images taken with the {\it Hubble Space
Telescope} ({\it HST}) make observing the host galaxy much easier.
The first, and to date one of the largest, systematic {\it HST}
studies of QSO hosts is by Bahcall, Kirhakos, \& Saxe~(1997), who
study 20 of the most luminous nearby QSOs.  They can discern the
morphology of the hosts, and they discover that, while radio-loud QSOs
are found only in ellipticals or interacting systems, radio-quiet QSOs
can be in ellipticals, spirals or interacting systems.  They also find
that QSO hosts do not follow a Schechter~(1976) luminosity function
and are instead found at systematically high luminosities.  More
recent studies strengthen and expand on these results.  McLure et
al.~(1999) confirm that QSO hosts are generally luminous, and also
determine that, even for radio-quiet QSOs, the hosts are often
ellipticals or bulge-dominated.  Furthermore, elliptical hosts appear
to follow the same luminosity-surface brightness relation as field
elliptical galaxies (Hamabe \& Kormendy~1987).  Other recent studies
include Nolan et al.~(2001), who discuss QSO host ages, and Kukula et
al.~(2001), who study host evolution from redshifts of $z \approx 2$
to the local universe.

In this paper we focus on the luminosity distribution of a large
sample of QSO hosts observed with the Wide Field and Planetary
Camera 2 (WFPC2) aboard {\it HST} (\S\ref{sec:sample}).  We
have collected and reanalyzed wide-band archival images of 71
QSOs with $ M_V \leq -23$ mag (total
nuclear $+$ host light) and redshifts $ 0.06 \leq z \leq 0.46 $.  We
have taken an inclusive approach in our sample selection, imposing no
additional selection criteria on the QSOs besides those of total
absolute magnitude and redshift, while some of the previous work on
QSO hosts has focused on specific classes of QSOs: radio loud (Lehnert
et al.~1999), intrinsically very bright (Bahcall et al.~1997), and so
on.  For each we have subtracted the nuclear light component using
two-dimensional image fits and have derived the luminosity and size of
the underlying host galaxy by fitting both an $ r^{1/4} $ and an
exponential light profile (\S\ref{sec:measurements}).  Given the total
number of objects considered, more than triple that of previous
studies, we can effectively sample the general QSO population for
redshifts $ z \leq 0.46 $, and derive a global luminosity function for
their host galaxies (\S\ref{sec:lumfun}) that is not grossly affected
by selection criteria.  This luminosity function is compared with that
of normal galaxies, and selection effects/biases and other issues are
discussed (\S\ref{sec:discussion}). Conclusions are then summarized
(\S\ref{sec:conclusions}).

Throughout this paper, we adopt a Friedman cosmology with $H_0 = 50$
km s$^{-1}$ Mpc$^{-1}$, $q_0 = 0.5$, and $\Lambda=0$.  We have
converted the results of other researchers to this cosmology when
comparing our results to theirs.

We confirm previous results that host galaxies of QSOs are
significantly more luminous than typical luminous $L^*_V$ galaxies,
where $L^*_V$ is the ``knee'' in the Schechter (1976) luminosity
function.  We also consider the relationship between host morphology
and QSO radio loudness.  Spiral hosts are, on average, nearly as
luminous as elliptical hosts, while hosts of radio-loud QSOs are, on
average, about 0.5 magnitudes brighter than hosts of radio-quiet QSOs.
Subject to systematic uncertainties in normalization procedures, we
find that the combined low-redshift QSO host luminosity function has a
very different shape from that of normal galaxies but that they
coincide at the highest luminosities.  In approximate terms, at
redshift $z\approx0.26$ the ratio of the number of these QSO hosts to
the number of normal galaxies of luminosity $L_V$ is $\mathcal{R}
\approx (L_V/{11.48L^*_V})^{2.46}$, where $L^*_V$ corresponds to
$M_V^*=-22.35$.

\section{SAMPLE SELECTION}\label{sec:sample}

Our sample includes 71 QSOs with total magnitudes $ M_V \leq -23 $ and
redshifts $0.06 \leq z \leq 0.46$ with available WFPC2 observations in
wide-band filters.  The objects are listed in
Table~\ref{table:obslist}, along with the {\it HST} programs under
which they were obtained.  The median redshift of this sample is
$z=0.24$, close to the mid-point of the range considered.  The
two-thirds of the sample surrounding the median point falls within the
range $0.14 \leq z \leq 0.41$.  We have analyzed or reanalyzed {\it
HST} archival imaging observations in a systematic and uniform manner,
and the results are reported in Table~\ref{table:obslist}.
Comparisons are made between the magnitudes we determine and the
results of the original observers in \S\ref{sec:comparisons}.  The
absolute magnitude selection ($M_V \leq -23$) aims at including only
historically traditional QSOs.  However, since the selection is based
on the combined magnitude of the host and nucleus, lower-luminosity
nuclei, down to $ M_V \approx -19 $, are in fact included in our
sample.  We have not excluded these objects from our analysis since
they would be present in most magnitude-selected ground-based samples.

We limit our analysis to objects with $z \leq 0.46$, in order to
obtain a significant sample size and to ensure that the {\it HST}
resolution permits a reasonably reliable separation between a host and
nuclear component.  At $ z \approx 0.4 $, a typical host with a
half-light radius of 9 kpc has an apparent radius of $1\farcs4$, which
corresponds to 14 pixels in the Wide Field Camera (WFC) or 31 pixels
in the Planetary Camera (PC).  The light from a luminous host should
therefore be clearly separated from that of the nucleus.  In fact, the
host galaxy cannot be convincingly detected in only one of the 71 QSOs
in our sample; this one is listed as such in
Table~\ref{table:obslist}.

Radio-loudness data are collected primarily from Brinkmann et
al.~(1997) and Yuan et al.~(1998), both of whom use a loudness
criterion that classifies an object as radio-loud if it has a
radio-to-optical flux density ratio in excess of 10.  Radio-loudness
data for the remaining objects come from a variety of sources, with
extensive use made of the NASA Extragalactic Database (NED).

\section{MEASUREMENT TECHNIQUE}\label{sec:measurements}

Even at {\it HST} resolution, the light of the unresolved nuclear central
source significantly affects the extended light distribution of the host
galaxy.  A careful subtraction of the central point source is needed in
order to measure the properties of the host accurately.  The following
is a brief description of our technique, which is largely similar to
that of Remy et al.~(1997).

Because of the complex structure of the {\it HST} WFPC2 Point Spread
Function (PSF), our analysis procedure has three principal steps: (1)
A model of the PSF is fitted to the central point source, in order to
determine its subpixel position and the telescope focus, which affects
the shape of the PSF.  (2) The PSF and a galaxy model are
simultaneously fitted to the entire image to distinguish the nuclear
and host components.  (3) The nuclear magnitude is determined from the
PSF model.  Then the fitted PSF is subtracted, and the magnitude of
the host is determined from the residual light.

The fitting of a model PSF, as opposed to an observed PSF, is dictated
by both opportunity and quality considerations.  Since we rely on
archival data, in most cases we do not have a PSF observation taken at
the same time as the QSO image.  The PSF in WFPC2 varies with time as
a consequence of short and long term changes in the telescope focus.
Thus, using a PSF observed at other times does not generally yield a
good subtraction of the nuclear light.  Also, the undersampled nature
of WFPC2 images make PSF subtraction very difficult, unless both PSF
and image have been properly dithered.  Under these circumstances, a
cleaner PSF subtraction can be achieved by using a model PSF produced
by the TinyTim software (Krist \& Burrows~1995), provided both focus
and subpixel positions are explicitly fitted (Remy et al.~1997; Surdej
et al.~1997).  This also results in photometry that is comparable in
accuracy to using an observed PSF.

\subsection {First Step: Fitting the PSF}\label{sec:psffit}

The model PSF is constructed from a set of artificial PSFs, created using
the TinyTim software (Krist \& Hook~1999).  TinyTim uses a detailed model
of the telescope and camera optics, including the zonal errors in the
primary and secondary mirrors, to produce a good wavelength-dependent
approximation of the resulting PSF.  However, the PSF structure
changes significantly depending on both the telescope focus and on
exactly how the point source is centered with respect to the pixel grid.
The telescope focus changes with time due to ``breathing,'' which is the
thermal expansion and contraction of the spacecraft due to changes in its
attitude relative to the Sun.  Breathing typically changes the relative
positions of the primary and secondary mirrors by about $5 \micron$.

Therefore, we produce PSF models oversampled 11 times, {\it i.e.}, on
virtual pixels 11 times smaller in area than actual detector pixels,
and for focus positions that range from $ -10 \micron$ to $ +10
\micron$ in $ 1 \micron$ steps.  Each PSF is then aligned with various
offsets with respect to the true pixel grid and resampled to the
actual detector resolution, including the estimated pixel spread
function described in the TinyTim documentation.  The best fit to the
light distribution in the central few pixels identifies the subpixel
position and the estimated focus of the observation.  If the PSF is
not saturated, we can achieve a precision of $\approx 0.01$ pixels in
the central position and $\approx 1 \micron$ in the focus position,
for QSOs dominated by their nuclei.  During this procedure, the light
of the extended galaxy, which varies little over the scale of the PSF,
is treated as a constant background.

In some cases, several of the pixels at the core are overexposed and
saturated on the CCD.  These pixels provide no information and are
masked from all fits.  The pixels vertically adjacent to them are also
masked, because the CCD ``blooming'' effect could have altered those
pixels' intensity values.  Our technique works in the presence of
saturation, although the focus position is determined less accurately.
Most images have either no saturation or a small amount that does not
completely cover the PSF core.

Once the position and focus have been found, a PSF of angular size
large enough to cover the host image is created with these parameters,
and it is used in the subsequent analysis.

\subsection {Second Step: Distinguishing the QSO and 
Host Galaxy Light}\label{sec:distinguish}

A second two-dimensional fit distinguishes the light of the QSO from
that of the resolved host galaxy, simultaneously fitting both parts.
In this step, the model PSF's brightness is scaled to match the
QSO nuclear brightness, while a galaxy model is fitted to the host.
The host model accounts for ellipticity, position angle, brightness,
size, and a simple morphological classification based on radial profile.
We consider two surface brightness models, each of which is convolved
with the PSF: the de Vaucouleurs $ r^{1/4} $ law, $I(\tilde{r}) = I(0)
\exp{[-7.67(\tilde{r}/\tilde{r}_{1/2})^{1/4}]}$, which is typical of
elliptical galaxies, and the exponential law, $I(\tilde{r}) = I(0)
\exp{[-\tilde{r}/\tilde{r}_e]}$, typical of spirals.  Here, $\tilde{r}$
is the elliptical radius, $\tilde{r} = (x^2 + \alpha^2 y^2)^{1/2}$, where
$x$ and $y$ are aligned with the major and minor axes of the ellipse,
respectively, and $\alpha = a/b$, where $a$ is the semi-major axis and
$b$ is the semi-minor axis.  The half-light radius $ \tilde r_{1/2} $
is the elliptical radius enclosing half the total light as projected onto
the sky; for the exponential model, $ \tilde r_{1/2} = 1.68 \tilde r_e $.

In six cases, namely PG~0052$+$251, MRK~1014, PKS~0736$+$01, 3C~215,
LBQS~1222$+$1010 and PG~1402$+$261, the automated fitting procedure
does not produce a good match to the central point source, most often
because of complex host features at very small radii.  For these
cases, we manually subtract an increasingly luminous central point
source until the residuals are smooth.  Consequently, the resulting
nuclear and host magnitudes for these objects are somewhat subjective
and more uncertain.

\begin{figure}
\plotone{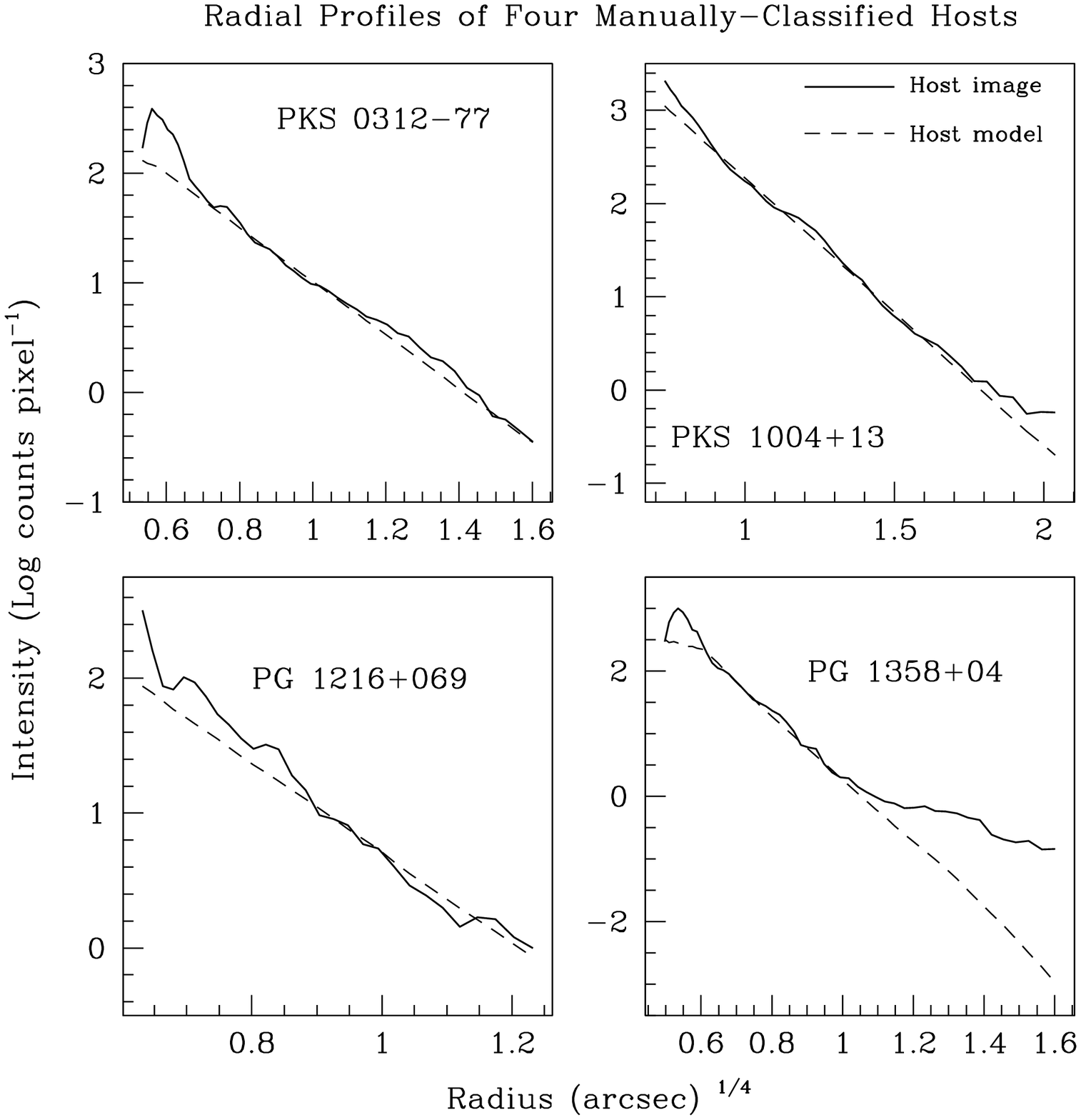}

\caption{Radial profiles of four manually--classified hosts.  These
are the four classified as spirals by the automated fit but whose
profiles more closely follow an $r^{1/4}$ law, as shown here.  The
PSFs and sky backgrounds have been subtracted to show the profiles of
the underlying hosts (solid lines).  The profiles of the fitted
elliptical host models (dashed lines) are shown for comparison.  The
straightness of the profiles shows that these hosts generally follow
the $r^{1/4}$ law.  For PG~1358+04, this is true for the bright, inner
regions.  It has a distinct second component at large radii that is
not modeled and is masked from the fit, but this only exists in
regions fainter than 1 count pixel$^{-1}$.  The tilt of PG~1216+069's
host profile relative to the model is due to the weighting scheme and
the imperfect fit of the bright PSF.  On the basis of these profiles,
all four of these hosts are classified as ellipticals.}

\label{fig:four-prof}
\end{figure}

For $\approx 90$\% of the objects, the host morphology is assigned
simply on the basis of the best-fitting (lowest $ \chi^2 $) model:
elliptical if the $ r^{1/4} $ model fits best, spiral if the
exponential model does.  We overrule the automatic classification in
seven cases.  Four hosts, those of PKS~0312$-$77, PKS~1004$+$13,
PG~1216$+$069, and PG~1358$+$04, yield a spiral classification after
the automated fit, but plots of their radial profiles
(Figure~\ref{fig:four-prof}) show them to follow an $r^{1/4}$ law more
closely overall [{\it i.e.}, plotting log(counts) vs. $r^{1/4}$ yields
a straight line], and they show no evidence of spiral arms.  They are
reclassified as ellipticals, and their $r^{1/4}$ models are used in
our subsequent analysis.  Three hosts, those of MRK~1014,
PG~1309$+$355 and MS~2159.5$-$5713, have their radial profiles outside
their central bulges well represented by the $r^{1/4}$ law, yet they
show clear evidence of spiral arms.  They are thus classified as
spiral, although their best-fitting $r^{1/4}$ models are used in the
analysis.  Note that we are able to identify late-type spiral
structure in hosts at redshifts as high as $z=0.4$.  Except for these
seven cases, we keep the morphological assignments determined by the
best-fitting models.

For spiral hosts with a visible bulge, we use masks to fit the bulge
and disk separately.  The bulge is fitted first, and its model is
subtracted from the entire image before the disk is fitted.  Bars, if
present, are masked out of the fit altogether, though they are used in
determining the host's total magnitude.  Based on visual inspection,
some hosts appear to have undergone recent, strong interactions that
have severely distorted their appearances from those of a normal
elliptical or spiral, and we have noted these in
Table~\ref{table:obslist}.  Two QSOs in this sample have a nucleus not
concentric with the main part of the host, 3C~48 and IR~0450-2958.
Both of these appear to have undergone severe interactions with other
galaxies and are left with disrupted hosts and offset nuclei.  There
are several other cases of hosts that have undergone interactions,
such as Q~0316-346, PG~1012+008, and PKS~2349-014, but these still
have a nucleus centered within the host.

\subsection{Third Step: Extracting the Magnitudes}\label{sec:extractmag}

Using the fitted parameters, we then subtract the properly scaled PSF
from the QSO image, leaving the host galaxy.  The magnitude of the
nucleus is measured directly from the scaled PSF model.  The light of
the PSF model is measured within an aperture of radius $0\farcs5$.  An
aperture correction of 0.10 magnitudes is subtracted (Voit 1997), and
the result is used for the nuclear apparent magnitude.  The host
magnitude is measured from the PSF-subtracted image, within an
aperture large enough to encompass the visible extent of the host.
Outside the aperture, we extrapolate the {\it host model} to a radius
of infinity and add this contribution to the light contained within
the aperture, yielding the apparent magnitude of the host galaxy.  The
measurements are not based on the models alone because the host
profiles often deviate from strict $r^{1/4}$ or exponential laws,
creating noticeable differences between magnitudes derived from the
model alone and magnitudes obtained in the way described above.  For
some objects, we find differences between our host magnitudes and
those of other researchers that may be due to their having derived the
magnitudes from the host model alone (see \S\ref{sec:compmethod} and
\S\ref{sec:comparisons}).

With the exception of MRC~0022$-$297 (which was observed with the
F336W filter), the observations were made in WFPC2 F606W or redder
filters.  The measured apparent magnitudes of the nuclei and hosts are
transformed to rest frame $V$.  They are then converted to absolute
magnitudes in our adopted cosmology and reported in
Table~\ref{table:obslist}.

In calculating the absolute $V$ magnitudes, the apparent $V$
magnitudes are first obtained by applying a color correction.  Colors
for the nuclei are interpolated in redshift from Cristiani \&
Vio~(1990), who provide $\vr$ and $V-I$ as functions of the redshift,
$z$ (where $R$ corresponds closely to $F675W$ and $I$ to $F814W$).  We
obtain colors for $V-F606W$ and $V-F702W$ by treating these filters as
linear combinations of $V$, $R$, and $I$.  We calculate the
combinations using the IRAF {\sc synphot} package and a power-law
spectrum of the form $f_\nu \propto \nu^\alpha$, with $\alpha = 0.0$.
For the $F606W$ filter, we use $V = F606W + 0.25(\vr)$, and for the
$F702W$ filter, $V=F702W + 0.85(\vr) + 0.15(V-I)$.  For $R$-band images of
redshift $z < 0.1$ and $I$-band images of redshift $z < 0.2$,
Cristiani \& Vio (1990) have no color data.  We therefore take
$\vr=0.25$ and $V-I=0.43$, calculated using the above power-law
spectrum.

Galaxy colors are interpolated in redshift from Fukugita et
al.~(1995).  Because we do not classify the spiral galaxies into more
detailed morphologies, we average together the colors given for the S0
and all spiral categories in Fukugita et al.~(1995).  There are
noticeable differences in the $V-F814W$ colors of S0 and Scd galaxies
at higher redshifts.  At a redshift of $z=0.0$, the maximum deviation
from the average is $\approx 0.05$ mag, but this grows to $0.26$ mag
at a redshift of $z=0.5$. The other spiral colors have much smaller
differences.

Once we have the apparent $V$ magnitudes, the absolute $V$ magnitudes
are given by
\begin{equation} 
M_V = m_V - 45.396 - 5 \log (1+z - \sqrt{1+z}) - K(V) - A_V \mbox{ ,}
\end{equation}
where $K(V)$ is the $V$-band K-correction, and $A_V$ is the Galactic
reddening.  The individual absolute magnitudes here account for
Galactic extinction, using the data from Schlegel, Finkbeiner, \&
Davis~(1998) and interpolated by the Galactic extinction calculator on
the NASA Extragalactic Database \footnote{\tt nedwww.ipac.caltech.edu}
(NED).  The unweighted and weighted host luminosity distributions use
the extinction-corrected magnitudes, but the host luminosity function
uses the uncorrected magnitudes.  This is done to allow a direct
comparison with the luminosity function of Metcalfe et al.~(1998),
which does not account for Galactic extinction.  We interpolate the
nuclear K-corrections in redshift from the data of Cristiani \&
Vio~(1990).  For galaxy K-corrections, we use the data of Pence~(1976)
which assumes no intrinsic reddening in the host galaxies.
Pence~(1976) combines E and S0 morphologies into a single category and
subdivides spiral galaxies into multiple categories.  Following our
decision with spiral galaxy colors, we average the K-corrections for
all S-types, including S0.  At a redshift of $z=0.06$, the maximum
deviation from this average is $0.06$ mag, and at a redshift of
$z=0.46$, this increases to $0.49$ mag.

\subsection{Comparison with Other Methods}\label{sec:compmethod}

The analysis techniques used by others in studies with
criteria similar to our own are sometimes different in minor ways.  
For example, the method of Bahcall et al.~(1997) uses stellar PSFs 
taken at the
time of the observations, while we generally cannot.  They observe a
set of four stars for the PSFs, chosen to have colors similar to QSOs.
The PSF is subtracted by scaling it until the $\chi^2$ between it and
the QSO image is minimized.  The best-fitting of the four PSF stars is
used in each case.  Elliptical and spiral host models are then fitted to
the residual in an annular region $r > 1\farcs0$, avoiding the core of
the QSO.  Bahcall et al.~(1997) try one- and two-dimensional models,
and adopt the two-dimensional results in the end.

McLure et al.~(1999) also use stellar observations for the PSFs in
their analysis, employing two-point dithering to improve the sampling
(and therefore the subpixel centering), with the PSF stars being
chosen to match closely the typical $\bv$ colors of QSOs.  Their host
fitting technique is similar to ours, with the host and QSO being
fitted simultaneously.  They use a two-dimensional host model,
assuming either a strict $r^{1/4}$ or exponential law profile, and
varying the host model's size, luminosity, ellipticity and position
angle, as well as the nuclear luminosity.  Separately, they try the
technique of using a model with the radial profile having a variable
exponent, $\beta$.  A true exponential law would have $\beta = 1.00$,
and a de Vaucouleurs law would appear as $\beta = 0.25$.  They achieve
similar classifications using this technique, but their adopted
$M_{\mathrm{host}}$ values are based on the former method.

\section {THE LUMINOSITY FUNCTION OF QSO HOST GALAXIES}\label{sec:lumfun}

\subsection {Subclasses and the Unweighted Absolute Magnitude 
Distribution of QSO Hosts}\label{sec:unweighted}

With 70 detected QSO hosts in our sample, we are able to investigate
the properties of the host galaxy luminosity function quantitatively 
and consider issues related to host morphology and radio loudness.

As noted in \S\ref{sec:sample}, the host of one QSO is not
convincingly detected; this one, MRC~0022$-$197, is a radio-loud QSO
and is the only QSO observed in the F336W filter (approximately
Johnson $U$).  Importantly, this is also the QSO in our sample with
the faintest apparent magnitude, $m_V=19.0$, according to
V\'{e}ron-Cetty \& V\'{e}ron~(1998).  For one spiral
(MS~2159.5$-$5713), we have no radio information. Consequently,
MRC~0022$-$197 is excluded from all analyses, and MS~2159.5$-$5713 is
excluded from analyses requiring radio information.

\begin{figure}
\plotone{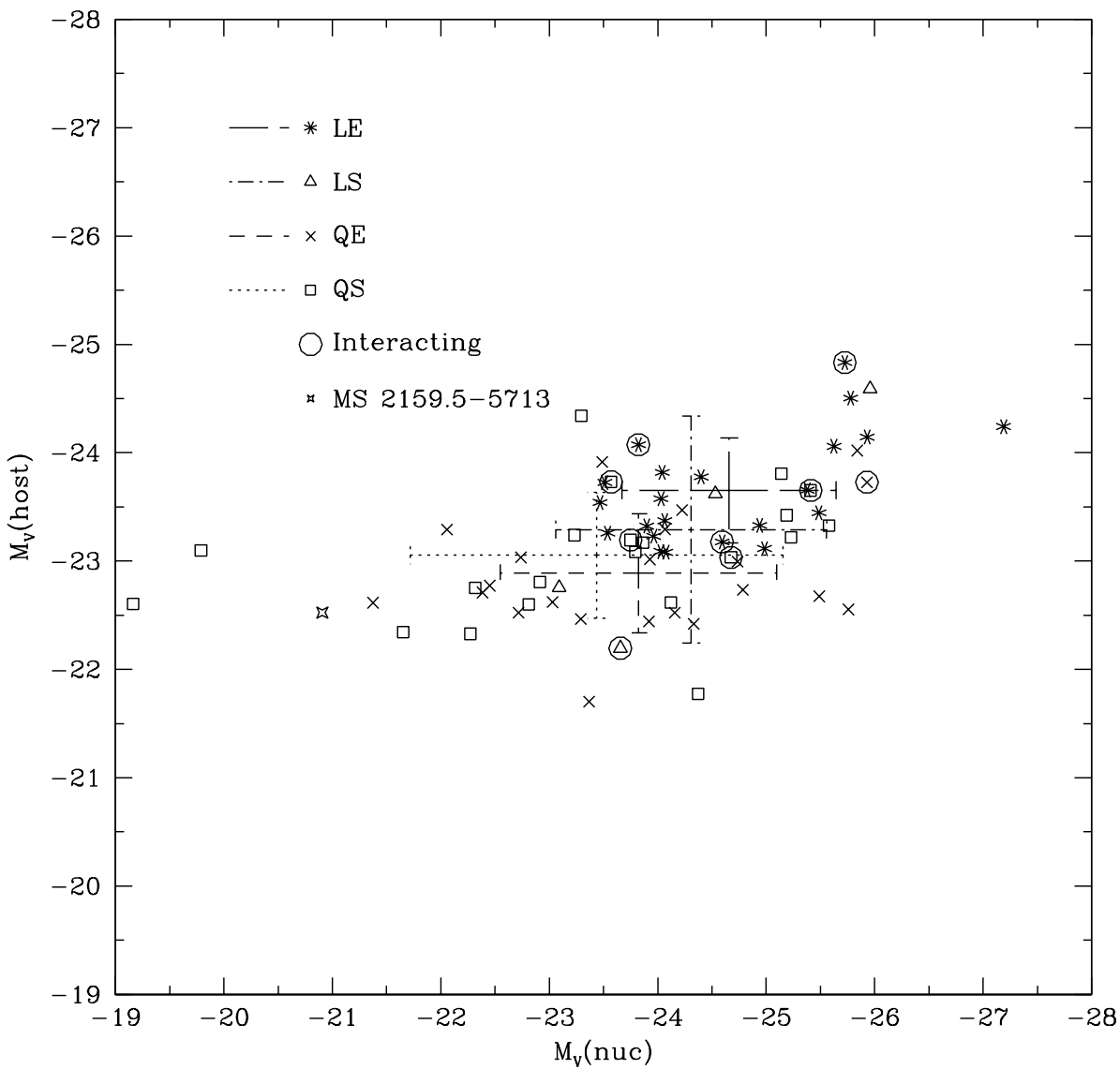}
\caption{Distribution of morphological and radio properties of the
sample with respect to host and nuclear luminosities.  ``LE'' refers
to radio-loud QSOs in ellipticals, ``QE'' refers to radio-quiet QSOs
in ellipticals, ``QS'' refers to radio-quiet QSOs in spirals, and
``LS'' refers to radio-loud QSOs in spirals.  The RMS error bars for
each subclass are overlaid, centered on the mean of each distribution
according to the legend.}
\label{fig:MvsM}
\end{figure}

It is of interest to consider whether the remaining elliptical and
spiral hosts in our sample are drawn from different parent
populations.  This might also be related to possible selection effects
and biases (\S\ref{sec:selectioneffects}).  Below we consider these
objects in terms of a binary classification yielding four subclasses.
The nuclear and host magnitudes for these objects, separated by
subclass, are plotted in Figure~\ref{fig:MvsM}. The overall absolute
magnitude distribution of the hosts in our sample is
shown in Figure~2. Our sample is divided
almost evenly into three subclasses: radio-loud QSOs with elliptical
hosts (designated ``LE,'' 22 objects), radio-quiet QSO with elliptical
hosts (designated ``QE,'' 22 objects), and radio-quiet QSOs with
spiral hosts (designated ``QS,'' 21 objects).  Membership in the
fourth subclass, radio-loud QSOs with spiral hosts (designated ``LS,''
4 objects), is rare. The total of these subsamples is 69 because
the lack of radio information on MS~2159.5$-$5713 excludes it.
First we discuss the rare LS subclass.

Two of the four radio-loud QSOs with spiral hosts, 3C~277.1 
and MC~1548+114A, are at
redshifts $z > 0.3$ and have little detail visible, but both appear to
have large tidal arms that may be responsible for the exponential
profile being the better model.  It is possible that they are simply
interacting cases and not normal spirals.  Additionally, 3C~277.1 is a
compact steep-spectrum quasar known to have bright, emission-line gas
aligned with the radio source, and its classification may have been
affected by this feature (De Vries et al.~1999).  A third object,
3C~351, appears to contain a complete ring surrounding an off-center
bulge, with the putative bulge following an $r^{1/4}$ radial profile.
We classify the complete host as a spiral on the basis of the ring
structure, although it could be another case of an interacting system.
The fourth, PG~1309+355 ($z=0.184$), has spiral arms but follows an $r^{1/4}$
profile.  Its unitless radio-to-optical flux density ratio is $\approx
18$ (Kellermann et al.~1989).  Since Kellermann et al.~(1989) classify
QSOs with radio-to-optical flux ratios $>10$ as radio-loud, it would
be considered radio-loud by that standard.  However, its
ratio does lie between the peaks of the radio-loud and radio-quiet
distributions.  Furthermore, its observed 6 cm flux is only $\approx54$ 
mJy, despite its low redshift.  Therefore, it, too, might be considered a
questionable case for a radio-loud spiral.

Thus, with the possible exceptions of PG~1309+355 and 3C~351, we 
confirm the result of Bahcall et al.~(1997) that radio-loud QSOs are
almost exclusively found in elliptical or interacting hosts, while
radio-quiet QSOs may be found in elliptical, spiral or interacting
systems.  The host absolute magnitudes of the radio-loud spirals were
found to lie in the range $-22.2 > M_V > -24.6$, spread across our
overall host absolute magnitude distribution.

\begin{figure}
\plotone{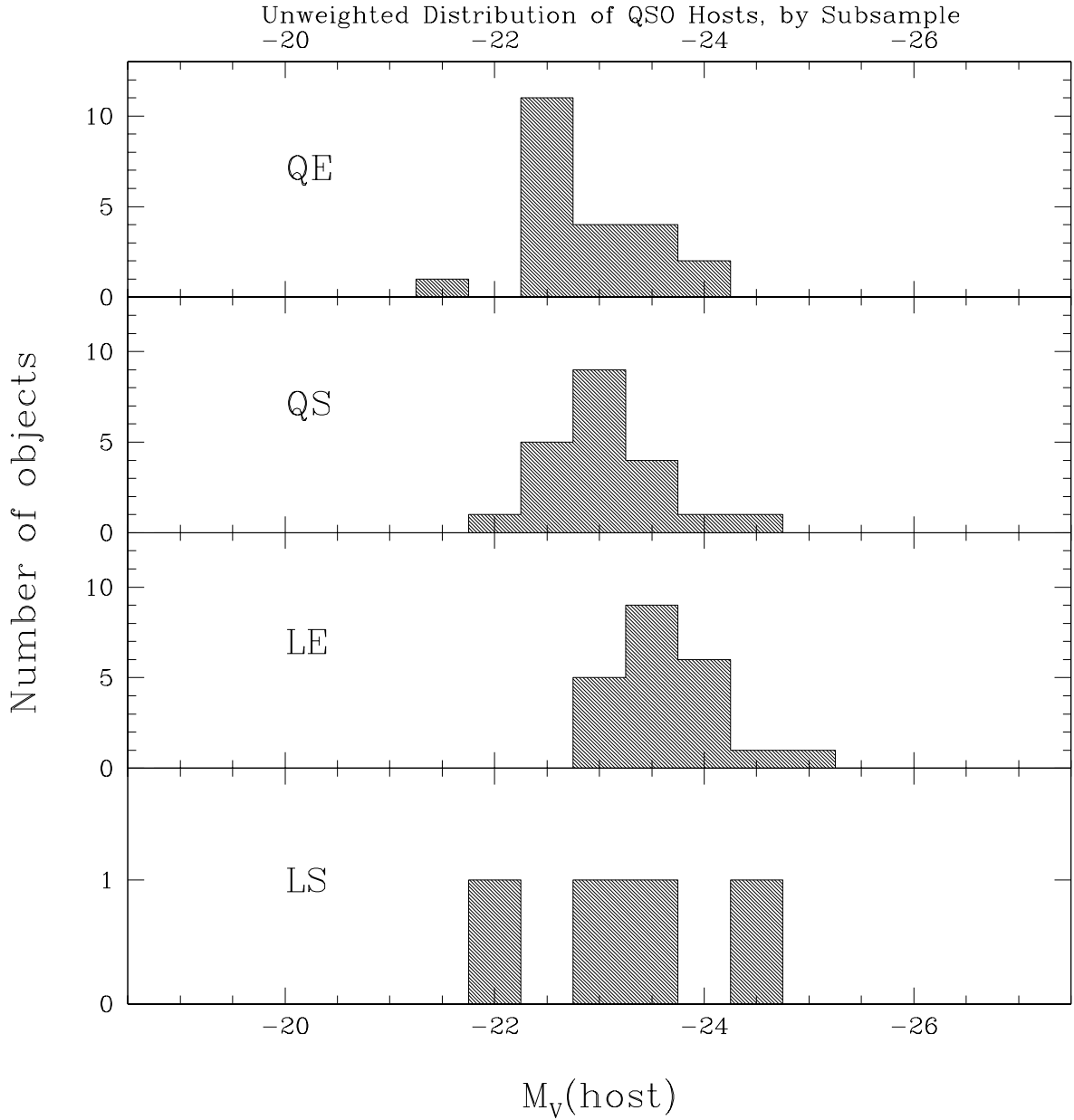}
\caption{Individual host distributions of radio-loud QSOs in
ellipticals (LE), radio-quiet QSOs in ellipticals (QE), radio-loud
QSOs in spirals (LS), and radio-quiet QSOs in spirals (QS).  Note that
the vertical axes cover the same range for all but the LS subsample.}
\label{fig:unwLF-sub}
\end{figure}

Table~\ref{table:medians} presents the median absolute magnitudes of
the various subsamples.  To check whether the host and nuclear
luminosities in each of the three major subclasses are consistent with
being drawn from similar parent populations, we apply two-sample
Kolmogorov-Smirnov (K-S) tests to each combination of subclasses. The
detailed results of this exercise are reported in
Table~\ref{table:kstests}.  The individual host magnitude
distributions are shown in Figure~\ref{fig:unwLF-sub}.  Comparison of
the LE and QE subclasses shows that their host luminosity
distributions differ at a significance $>99.9$\%, while their nuclear
luminosity distributions differ at a significance of $96.5$\%.
Formally, the LE subclass is more luminous than the QE subclass in
both cases.  For host magnitudes, the LE median is 0.8 mag brighter
than the QE median, but for nuclear magnitudes the LE median is only
about 0.2 mag brighter than that of the QE class.  Comparison of the
LE and QS subclasses shows that both their host and nuclear luminosity
distributions differ at a significance $>97$\%, with the LE subclass
again being the more luminous.  The LE median host magnitude is about
0.4 mag brighter than the QS median, and the LE median nuclear
magnitude is about 0.3 mag brighter than that of the QS class.  The
host luminosity distributions of the QE and QS subclasses are slightly
less distinct, differing at a significance of 88.1\%, but their
nuclear luminosity distributions are fairly compatible, differing only
at a significance of 9.2\%.

Additionally, we compare the host and nuclear luminosities of all
ellipticals (``E,'' 44 objects) to all spirals (``S,'' 26 objects),
and the host and nuclear luminosities of all radio-loud QSOs (``L,''
26 objects) to all radio-quiet QSOs (``Q,'' 43 objects). The results
are also shown in Table~\ref{table:kstests} and indicate that the
radio-loud and radio-quiet objects can be distinguished not only by
their nuclear luminosity distributions (98.7\% significance) but by
their host luminosity distributions as well ($>99.9$\% significance).
The hosts of radio-loud QSOs are typically half a magnitude more
luminous than their radio-quiet counterparts, and the radio-loud
nuclei are also noticeably brighter (0.3 mag) than the radio quiet
nuclei.  The differences between objects in ellipticals and spirals
are less significant, however.  Their nuclear luminosity distributions
are distinguishable, differing at a significance of 91.0\%, but their
host luminosity distributions differ only at 60.9\% level of
significance.  The ellipticals are more luminous than the spirals in
both cases.  The magnitude difference is fairly small ($\approx 0.2$
mag) in their median host magnitudes, but it is pronounced ($\approx
0.8$ mag) in their nuclear magnitudes.

\begin{figure}
\plotone{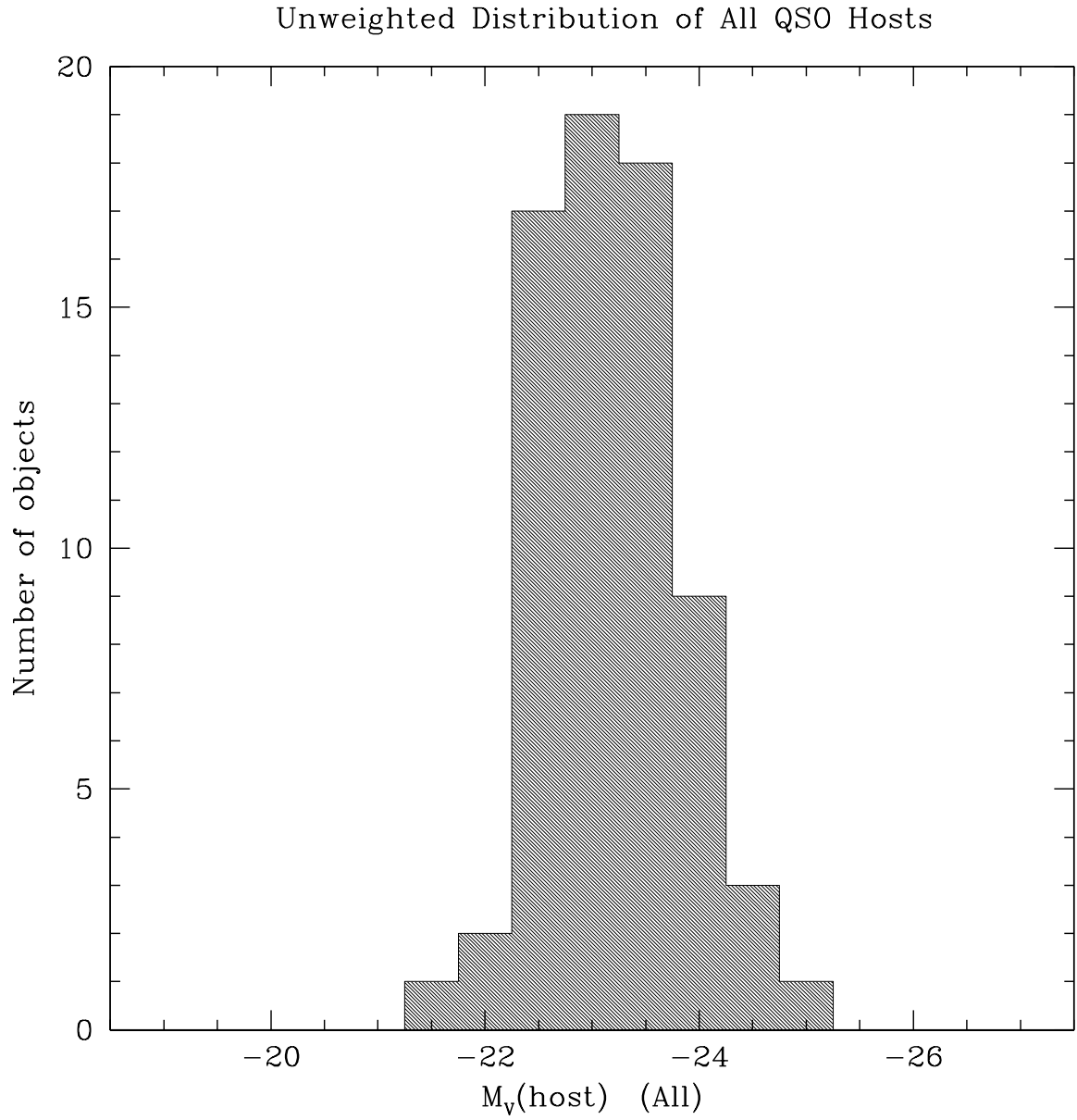}
\caption{Unweighted absolute magnitude distribution function of all
QSO hosts.  The narrow width of the distribution is evident.}
\label{fig:unwLF-all}
\end{figure}

From Figure~\ref{fig:unwLF-all} we see that the number distribution of
our complete sample of QSO hosts as a function of their absolute
magnitude.  This distribution is entirely contained within a range of
3.1 mag, from $M_V = -21.7$ to $-24.8$, with a median of $M_V =
-23.2$. It can be fitted by a Gaussian with a peak at $M_V = -23.1$
and a $1\sigma$ width of $0.67$ mag.  Note that since all but one of
the hosts are clearly detected, the lack of faint hosts is not due to
a failure to detect them.

\subsection {The Weighted Number Distribution of QSO Host 
Absolute Magnitudes}\label{sec:weightdist}

The 71 QSOs in our sample correspond to $\approx 7$\% of all known
QSOs within the magnitude range of our sample in the catalog of
V\'{e}ron-Cetty \& V\'{e}ron~(1998; hereafter VCV).  However, since
our selection of QSOs to include in this study is based on the
availability of {\it HST} observations, they may not adequately
represent the characteristics of all low-redshift QSOs.  In
particular, our sample may be systematically biased as a function of
apparent luminosity and redshift: nearer and brighter objects are more
accessible, and therefore more likely to be selected for study.  We
therefore apply a simple weighting technique to approximately correct
for redshift- and magnitude-dependent selection effects relative to
the VCV catalog. However, we note that any inherent biases in the VCV
catalog will not be removed.  This catalog is intended to be a
compilation of all known, published QSOs, so it has a mix of biases
from the various surveys that make up the catalog.  As a result, for
example, its ratio of the number of radio-loud to the number of
radio-quiet QSOs is not the true ratio, but we do not base our
conclusions on this information.  More subtle biases, such as those
involving limiting magnitudes, will remain.

Our procedure for weighting the distribution function to derive a
corrected or unbiased distribution function is based on a replacement
method as follows.  For each of the 982 QSOs in VCV within our
selected magnitude and redshift range, we pick a representative object
in our observed sample with approximately the same total (nuclear $+$
host) magnitude $M_{\mathrm{tot}}$ and redshift $z$.  The
representative object is chosen randomly with a Gaussian probability
distribution that depends on the difference in absolute magnitude and
redshift.  We choose a Gaussian width of $0.5$ mag in absolute
magnitude and $0.07$ in redshift.  These widths are chosen to ensure
that most catalog objects have several sample objects within about
1$\sigma$ in both magnitude and redshift; if the widths are too
narrow, regions of the $(M_{\mathrm{tot}},z)$ plane that contain few
sample objects would yield a luminosity function that depends too
heavily on those few objects.

\begin{figure}
\plotone{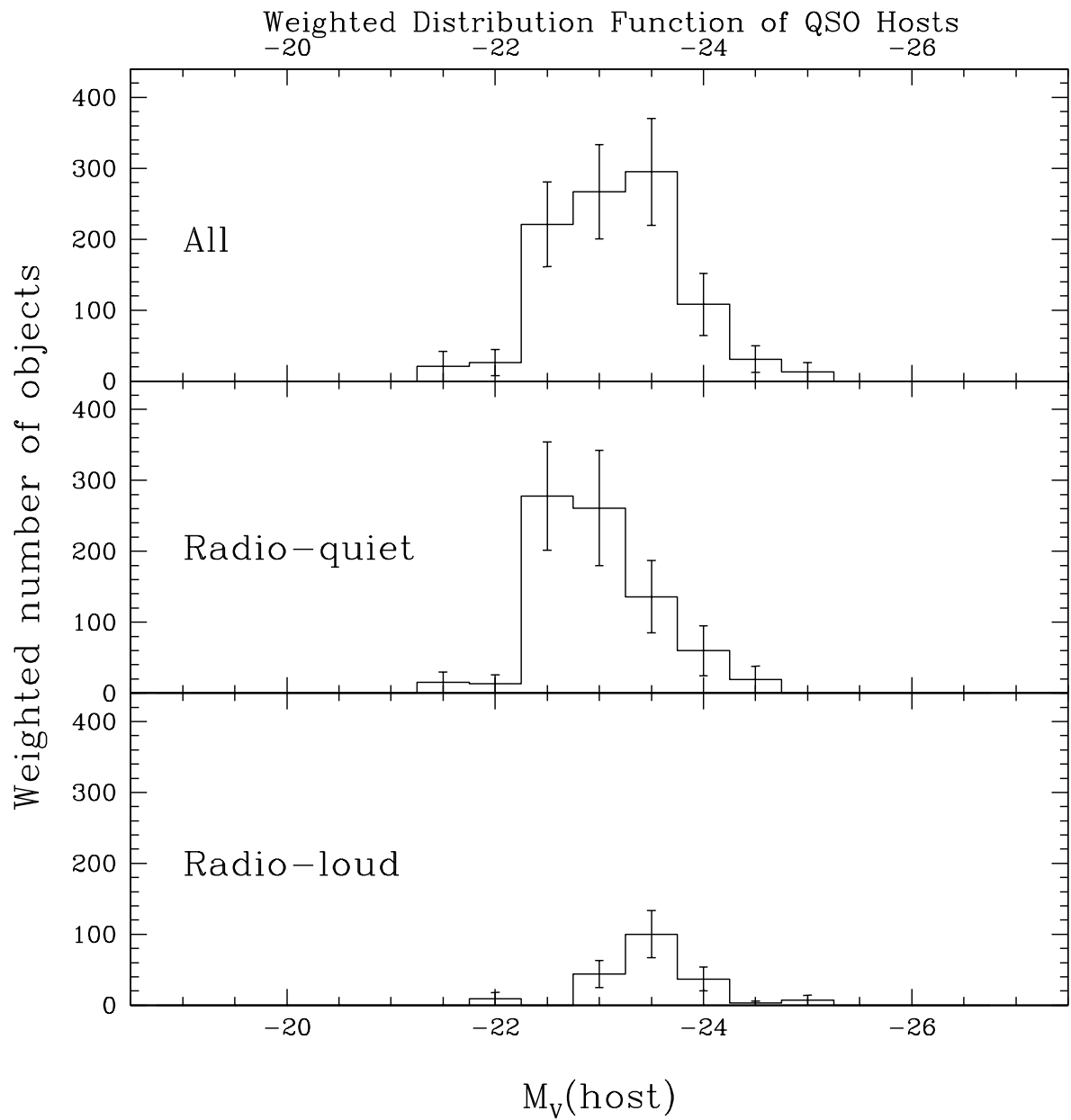}
\caption{Weighted absolute magnitude distribution functions of QSO
hosts.  The distribution for the complete sample is shown, as well as
the distributions for the radio-quiet and radio-loud subsamples.  Note
that the radio-quiet distribution has a higher population than the
total distribution in some bins, because they are calculated from
separate Monte Carlo runs.}
\label{fig:wLF}
\end{figure}

Each object, $i$, in our observed sample is then assigned a weight,
$w_i$, that is simply the number of times it is selected by the random
process.  The resulting weighted distribution function is shown as the
unshaded histogram in Figure~\ref{fig:wLF}.  The error bars reflect
the nominal counting error defined as $\sigma_{\mathrm{bin}} =
(\Sigma_i(w_i^2))^{1/2} $, where the sum is over all objects in the
bin.  Note that this error is an upper limit for the Poisson
uncertainty in the distribution function in that bin, in that it
assumes that the host magnitude is not correlated with redshift or
total magnitude; any correlation makes the assignment process less
random and therefore reduces the counting uncertainty.

The extinction-corrected, weighted distribution function has a shape
similar to that of the unweighted distribution, though with a narrower
peak, and its estimated median, $M_V \approx -23.23$, is almost
identical.  A Gaussian fit to the weighted distribution peaks at $M_V
= -23.1$ and has a $1\sigma$ width of $0.63$, so it peaks
approximately in the same location as the fit to the unweighted data
and has only a slightly narrower width.

Since the morphologies of VCV hosts are generally unknown, we cannot
weight the elliptical and spiral distributions separately.  However,
we can perform a simple weighting for radio loudness by using the
unitless radio-to-optical flux density ratio, $R_{\mathrm{ro}}$,
described by Kellermann et al.~(1989), calculated from the apparent
$V$ magnitudes and the 6 cm radio flux densities listed in VCV.  Since
these are relatively nearby QSOs, we assume those QSOs without radio
detections in VCV to have no 6 cm flux.  Kellermann et al.~(1989)
classify QSOs with $R_{\mathrm{ro}} > 10$ as radio-loud and those with
$R_{\mathrm{ro}} < 1$ as radio-quiet.  In keeping with Brinkmann et
al.~(1997) and Yuan et al.~(1998), we choose $R_{\mathrm{ro}} = 10$ as
a strict dividing line between radio-loud and radio-quiet.  Under
these assumptions, we derive a weighted distribution function of QSO
hosts, separated in terms of hosts of radio-loud and radio-quiet QSOs.
These individual weighted distributions are also shown in
Figure~\ref{fig:wLF}.  The shapes of the radio-loud and radio-quiet
distributions are fairly similar between the weighted and unweighted
versions.  Note that the weighted radio-quiet distribution rises
higher than the overall weighted distribution in some bins.  This is
because they are calculated from separate Monte Carlo runs.

\subsection {The Luminosity Function of QSO Host Galaxies}\label{sec:lf}

%Boyle's 6000 come from their 2QZ sample plus the LBQS and PG samples.

We use the QSO luminosity function of Boyle et al.~(2000) to derive a
normalization for our weighted host distribution function, turning it
into a QSO host galaxy luminosity function.  Boyle et al.~(2000) have
analyzed a ground-based sample of over 6000 QSOs to derive a QSO
luminosity function for the total (nuclear + host) light. They
parameterize the luminosity function in terms of a two-power-law
function,
\begin{equation} \Phi(M_B,z)=\Phi^*_{\mathrm{Boyle}} / 
\left \{ 
10^{-0.4(\alpha_{\mathrm{Boyle}}+1)[M_B-M^*_B(z)]} + 
10^{-0.4(\beta_{\mathrm{Boyle}}+1)[M_B-M^*_B(z)]}
\right \}
\mbox{ ,}
\end{equation}
and use a polynomial function for the evolution of $M^*_B(z)$ in
redshift, $M^*_B(z)=M^*_B(0) - 2.5(k_1z+k_2z^2)$, where
$\alpha_{\mathrm{Boyle}}=3.60$, $\beta_{\mathrm{Boyle}}=1.77$,
$M^*_B(0)=-22.39$, $k_1=1.31$, $k_2=-0.25$, and
$\Phi^*_{\mathrm{Boyle}}=6.8\times10^{-7}$ objects Mpc$^{-3}$
mag$^{-1}$.  Their data are limited to redshifts of $z \geq 0.35$, and
at the low-redshift end the data do not span a large range in total
absolute magnitude.  Thus, we restrict our consideration to redshifts
of $0.35 \leq z \leq 0.46$, where our sample overlaps with theirs, and
to total absolute magnitudes of $-23.00 \geq M_V\mathrm{(total)} \geq
-24.61$, extending no more than one magnitude brighter than
$M^*_B(z=0.405)$.  We use $z=0.405$ in the function since it is the
average of the range we consider.  We note parenthetically that 22 of
the QSOs in our sample lie within this redshift range and that 12 of
those also lie within the above total magnitude range.  Since the
total absolute magnitudes of the QSOs in the survey of Boyle et
al.~(2000) are likely dominated by nuclear luminosity, we assume $\bv
\approx 0.0$.

\begin{figure}
\plotone{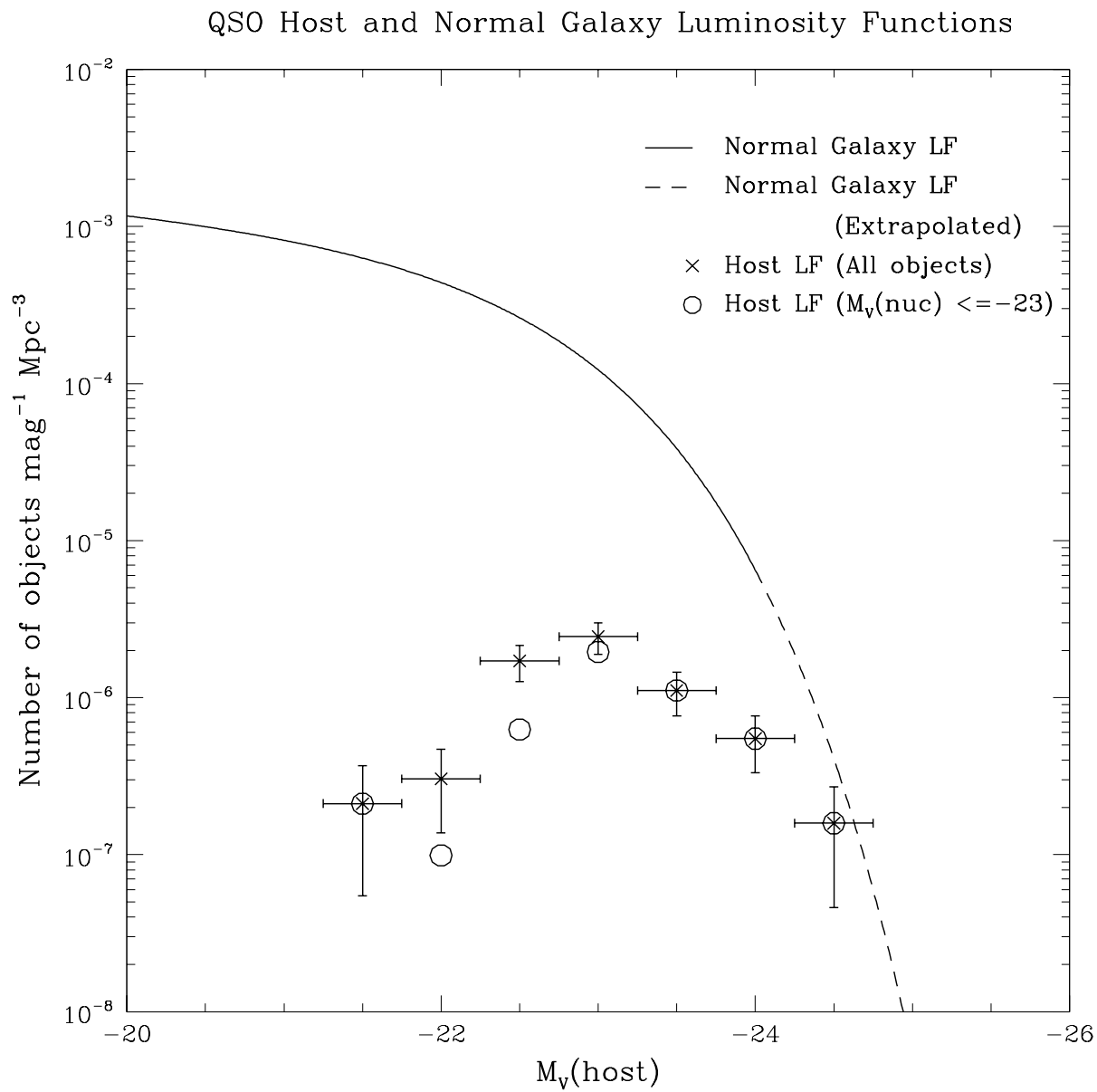}
\caption{Combined luminosity function of QSO host galaxies for our
sample compared to the normal galaxy luminosity function of Metcalfe
et al. (1998). The dashed line shows the extrapolated region of the
normal galaxy luminosity function.  Crosses show the derived
luminosity function for the entire sample, while open circles show the
derived luminosity function for QSOs with nuclear magnitudes brighter
than $M_V=-23$.}
\label{fig:complf}
\end{figure}

Integrating the two-power-law function over the range $-23.0 \geq
M_V\mathrm{(total)} \geq -24.61$, with $z=0.405$, we find $7.5 \times
10^{-7}$ QSOs Mpc$^{-3}$.  Over the same total absolute magnitude
interval, with $0.35 \leq z \leq 0.46$, there are $228$ objects in the
VCV catalog.  Dividing the integrated function by 228, we obtain a
normalization factor of $3.3 \times 10^{-9}$ Mpc$^{-3}$.  Multiplying
our weighted host distribution by this normalization factor and by a
factor of $2$ to account for our $0.5$ magnitude bin width converts
our distribution into a QSO host luminosity function in units of QSO
hosts Mpc$^{-3}$ mag$^{-1}$.  This QSO host luminosity function is
shown in Figure~\ref{fig:complf}. We note that in
Figure~\ref{fig:complf} we also show how removing objects with nuclear
luminosities fainter than $M_V=-23$ affects the derived luminosity
function.

\section {DISCUSSION}\label{sec:discussion}

Here we elaborate on some of the results of this work. A more thorough
discussion will be made elsewhere when we consider the other properties 
of the sample in detail.

\subsection {Comparison of the QSO Host and Normal Galaxy Luminosity 
Functions}\label{sec:normgal}

To compare our QSO host galaxy luminosity function (\S\ref{sec:lf})
with that of normal galaxies, we use the normal galaxy luminosity
function of Metcalfe et al.~(1998). The Schechter~(1976) luminosity
function parameters that describe their V-band luminosity function in
our cosmology are $\alpha=-1.2$, $M^*_V = -22.35$, and $\Phi^* =
8.5\times10^{-4}$ Mpc$^{-3}$.  The host luminosity function lies below
that of normal galaxies, as shown.  The most relevant uncertainty in
the normal galaxy luminosity function is at the bright end, where it
is less well constrained due to the dearth of luminous galaxies in
surveys.  The least luminous QSO host used in our analysis is
relatively luminous, with $M_V = -21.7$, and the median of our QSO
host luminosity function is at $M_V \approx -22.95$, twice the
luminosity of the $M^*_V$ ``knee'' of the normal galaxy luminosity
function.  The normal galaxy luminosity function is also shown in
Figure~\ref{fig:complf}, and it is evident that it has a very
different shape from that of the QSO luminosity function.

The brightest luminosity bin in the Metcalfe et al.~(1998) data
extends up to $M_V \gtrsim -24.0$, and their Schechter function is
extrapolated to brighter magnitudes, as we have marked with the dashed
line in Figure~\ref{fig:complf}.  In fact, we searched for published
accounts of any normal galaxies of luminosity $M_V < -24.0$ with
little success.  The Sloan Digital Sky Survey has recently released a
preliminary galaxy luminosity function (Blanton et al.~2001), in which
the highest-luminosity bin also extends up to $M_V \gtrsim -24.0$.  We
also searched the literature on Brightest Cluster Galaxies (BCG).  In
the sample of BCGs in Postman \& Lauer~(1995), the most luminous has
$M_V \approx -23.7$, although Disney et al.~(1995) refer in passing to
BCG luminosities as $M_V \approx -24.5$.  It may be that the $M_V <
-24.0$ region is dominated by QSO hosts.  We conclude that the
luminosity function of QSO hosts differs greatly in shape from that of
normal galaxies but that they are coincident in the highest luminosity
bin, if the normal galaxy luminosity function can be extrapolated that
far.

\begin{figure}
\plotone{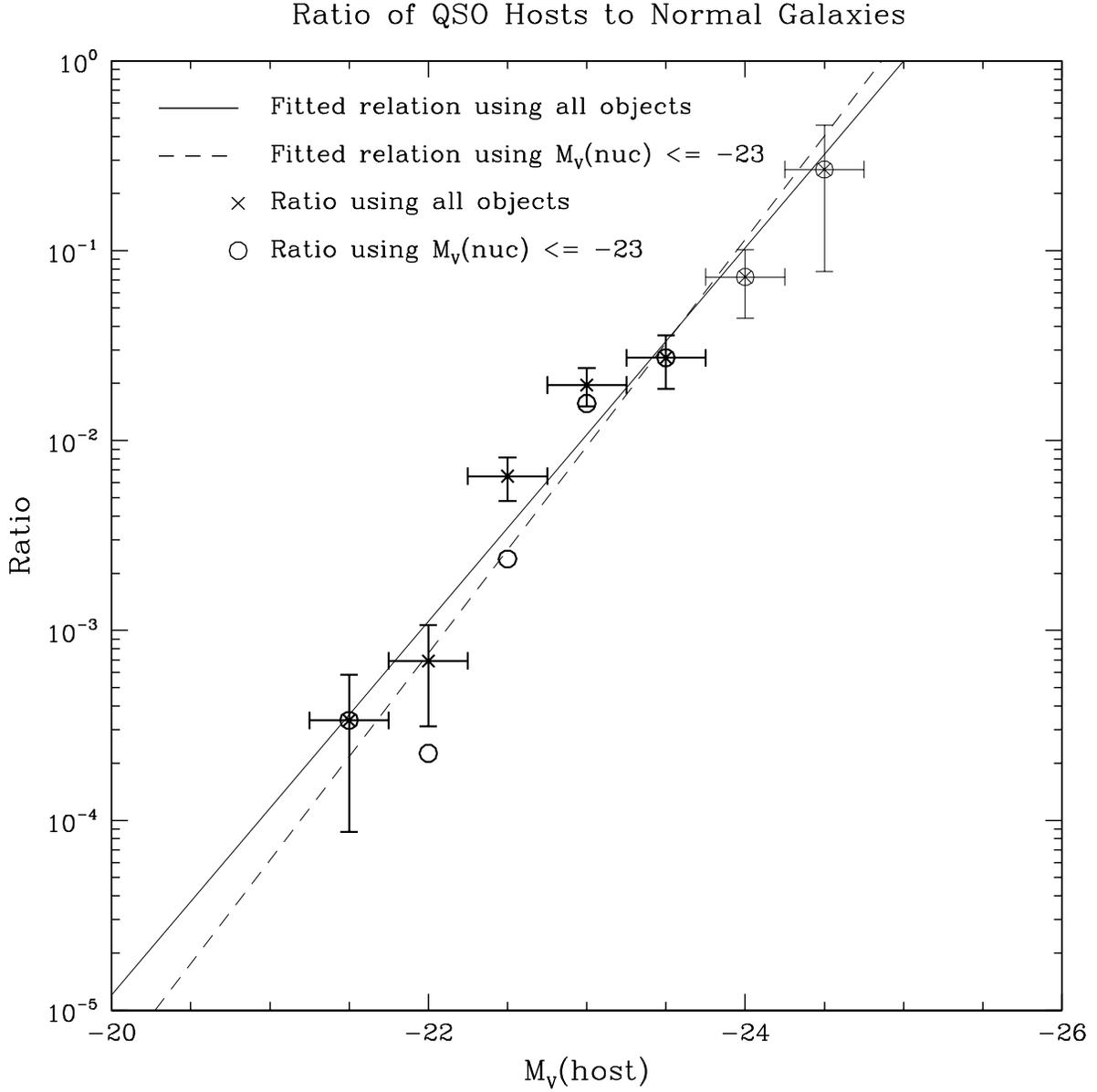}
\caption{Low-redshift QSO host galaxy luminosity function divided by
the normal local galaxy luminosity function, yielding the probability
that a galaxy of absolute magnitude $M_V$ will host a QSO. Crosses
show this result for the entire sample, while open circles show this
result for QSOs with nuclear magnitudes brighter than $M_V=-23$.  The
solid line shown in the figure is the fit specified in the text for
all hosts, while the dashed line is the fit for QSOs with
$M_V\mathrm{(nuc)}=-23$.  The upper two points, drawn with thin lines,
are in the extrapolated region of the normal galaxy luminosity
function.}
\label{fig:ratio}
\end{figure}

We can use the luminosity functions to estimate the ratio of the
number of low-redshift QSO hosts to the number of normal galaxies as a
function of absolute magnitude.  The normal galaxy luminosity function
is integrated over our 0.5--mag bins and then compared to the binned
host data.  The ratio of the number of hosts to the number of normal
galaxies is shown in Figure~\ref{fig:ratio}, along with a
parameterization of the results.  The relationship can be expressed as
$\mathcal{R} = (L_V/{11.48L^*_V})^{2.46}$, up to $M_V \approx -25.0$,
where $\mathcal{R}$ is the ratio of the number of hosts to the number
of normal galaxies, $L_V$ is the $V$-band luminosity, and $L^*_V$
corresponds to $M^*_V=-22.35$.  The points at $M_V=-24.0$ and $-24.5$
are drawn with thin lines to indicate that they depend on the
extrapolation of the normal galaxy luminosity function, although they
are included in the fit.  We note that the accuracy of these results
is subject to the inherent systematic uncertainties in normalization
procedures for both the local galaxy luminosity function and our
low-redshift QSO host galaxy luminosity function.  Normalizations of
normal-galaxy luminosity functions can differ by a factor of 2 from
one study to another.

The conclusions drawn here about the comparison of the QSO and normal
galaxy luminosity functions are roughly similar to those reached by
Smith et al.~(1986) in their ground-based study of the hosts of QSOs
and lower-luminosity AGN.  The current data are of course better than
what was available to them, so we believe that the conclusions are now
considerably stronger.  As suggested by these authors, this ratio
essentially represents an empirical parameterization of the
probability that a galaxy with luminosity $L_V$ will host a QSO. It
applies in the redshift interval studied here, $z\approx 0.06-0.46$.

\subsection {Radio-loud and Radio-quiet QSOs with 
Elliptical Hosts}\label{sec:similar}

As described in \S\ref{sec:unweighted} and shown in
Figure~\ref{fig:MvsM} and Figure~\ref{fig:unwLF-sub}, the QE subclass
tends to have dimmer host and nuclear luminosities than the LE
subclass, but there is a great deal of overlap.  Within this overlap,
there must be some property, other than host luminosity, that affects the
amount or nature of the fuel available to the central engine for radio
emission.  The possible causes of this effect should be investigated
more closely by using the members of these two subclasses of
elliptical hosts to examine other properties of these QSOs, including
environmental clues.

\subsection {Selection Effects and Biases}\label{sec:selectioneffects}

Images in our sample are of heterogeneous quality, because the
original observers have chosen a variety of filters and exposure times
for objects with disparate properties.  The best-exposed images, such
as Q~1402$+$436 and PG~1444$+$407, average up to 15000 e pixel$^{-1}$
within the half-light radius, for a S/N up to 150 per resolution
element (the full-width at half-maximum of the PSF).  The worst cases,
such as 3C~93 and LBQS~0020$+$0018, have as little as 200 e
pixel$^{-1}$, for a S/N of 25 per resolution element.

\begin{figure}
\plotone{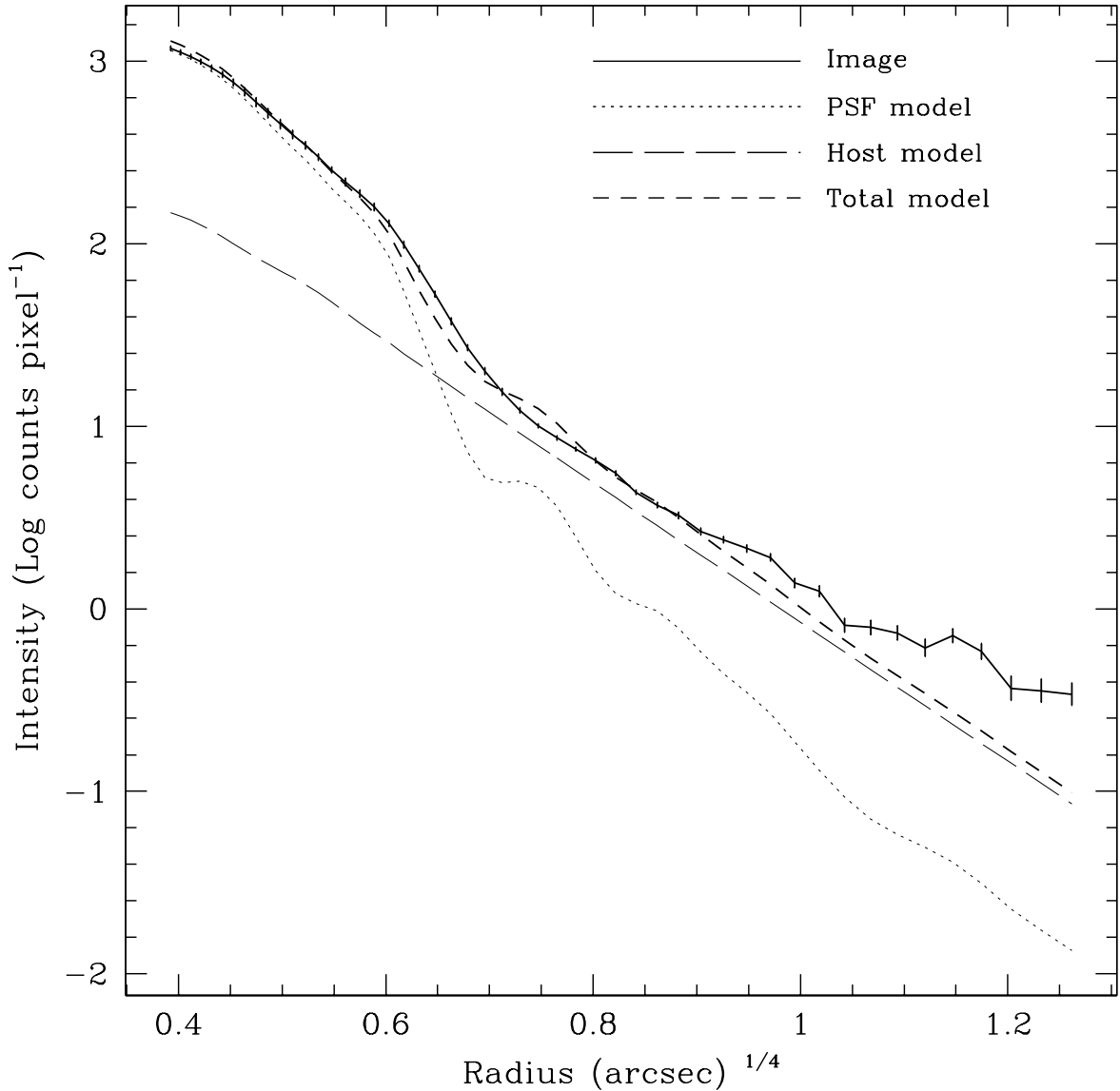}
\caption{Radial profiles of 3C~93 (solid line) and model.  Error bars
on the image profile represent uncertainties in the elliptical
isophotes used to plot the profile.  The model PSF is represented by
the dotted line, the model host by the long-dashed line, and the total
model by the short-dashed line.  Due to the low S/N, the model is not
a perfect match to the image profile.  Note the deviation of the host
from a strict $r^{1/4}$-law profile at large radii.  Though this
feature is not fitted by the model, the extra light is included in the
host magnitude.}
\label{fig:profile-3c93}
\end{figure}

\begin{figure}
\plotone{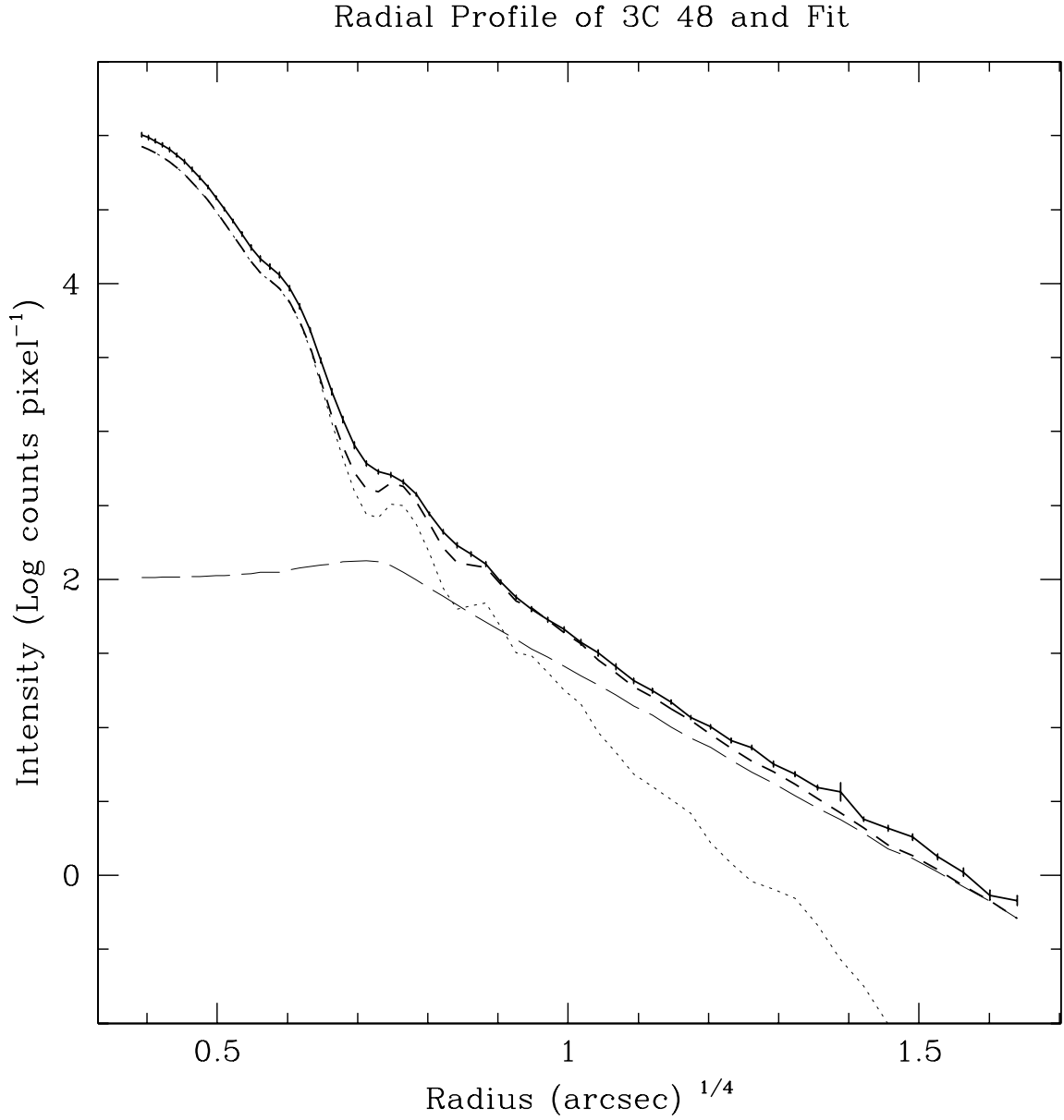}
\caption{Radial profiles of 3C~48 and model.  The higher S/N makes a
smoother profile than in the case of 3C~93, but the irregular host
shape causes the model to underestimate the light at small radii
slightly.  The turnover of the host profile is due to its center being
offset from the QSO nucleus by $0\farcs25$.  The line styles and axes
are the same as for Figure~\ref{fig:profile-3c93}.}
\label{fig:profile-3c48}
\end{figure}

\begin{figure}
\plotone{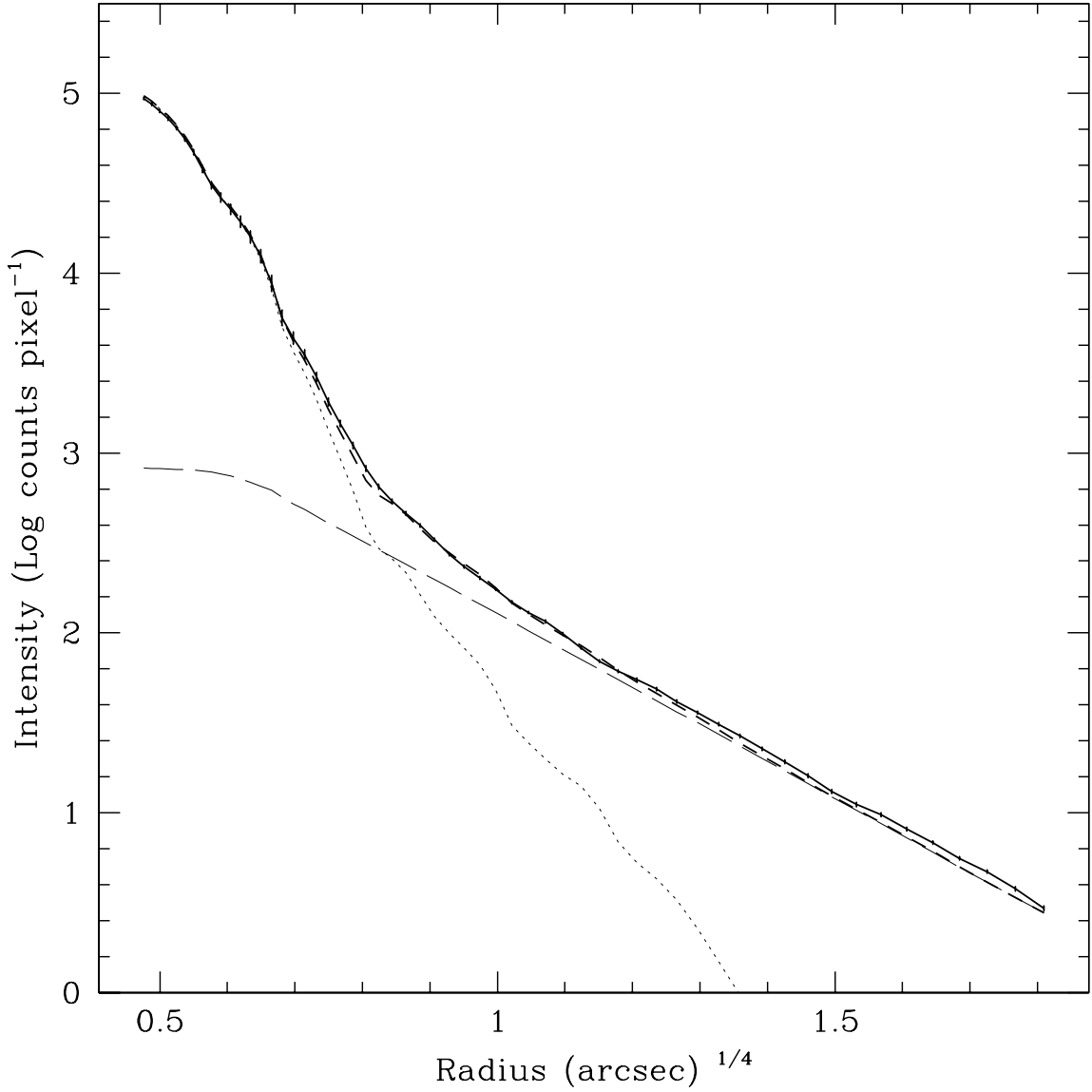}
\caption{Radial profiles of PKS~2349$-$014 and model.  Despite the
irregularity of the host morphology, a good fit to the profile is
obtained.  The line styles and axes are the same as for
Figure~\ref{fig:profile-3c93}.}
\label{fig:profile-pks2349}
\end{figure}

\begin{figure}
\plotone{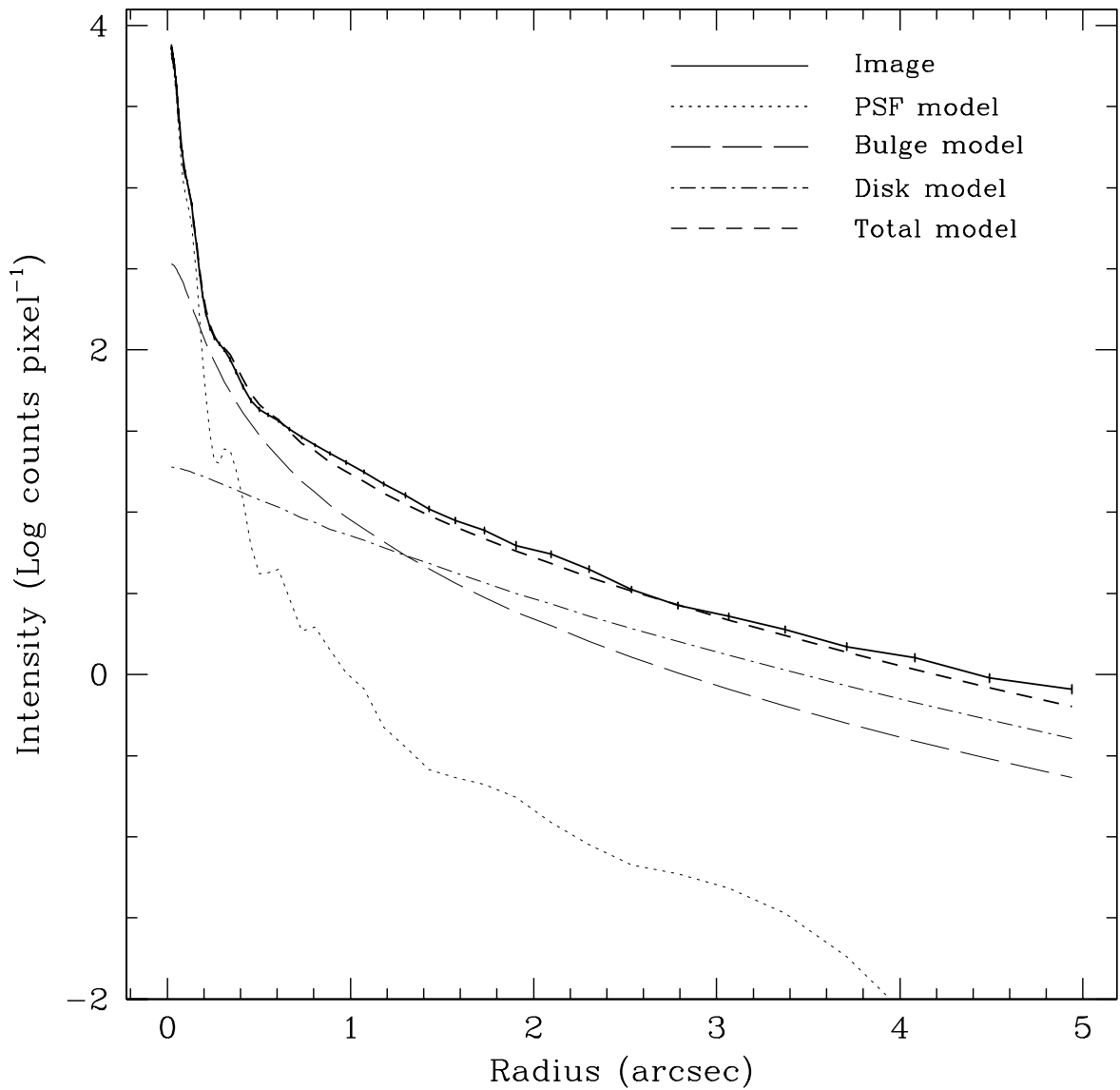}
\caption{Radial profiles of MS~1059.0$+$7302 (solid line) and model.
Note that line styles and axes differ from the previous figures.  The
dotted line represents the PSF model, the long-dashed line the bulge
model, the dot-dashed line the disk model, and the short-dashed line
the total model.  This QSO has a spiral host with a bulge and a disk.
The profile is not perfectly smooth but is fairly well modeled.  The
host model is slightly fainter than the image profile at large radii,
but the excess light is included in the host magnitude.}
\label{fig:profile-ms1059}
\end{figure}

Adequate model fits can be obtained even in images with low S/N, as
shown by the radial profile of 3C~93 (Figure~\ref{fig:profile-3c93}).
While noise does give its profile a ragged appearance, the model is
not a bad fit overall, and the small discrepancy between model fit and
measured profile is not reflected in the object's computed magnitude,
which is based primarily on the actual counts rather than on the model
fit (\S\ref{sec:extractmag}).  Some of the most irregularly-shaped
hosts can give rise to systematic differences between the image
profile and the model, as in the example of 3C~48
(Figure~\ref{fig:profile-3c48}).  In this case, the host's shape and
position, with the active nucleus offset from its center, requires
masking irregular parts of the host to allow the fit to converge; this
results in the systematic errors at small radii shown in the figure.
But others of these irregular hosts, such as PKS~2349$-$014
(Figure~\ref{fig:profile-pks2349}), have excellent fits, without
significant systematic differences in their profiles.  Finally,
Figure~\ref{fig:profile-ms1059} shows an example of a spiral host with
both the disk and bulge modeled.

A bright QSO may be expected to hide a dim host, and the more distant
it is, the harder the host will be to detect.  However, we are able to
see hosts that are 3.3 magnitudes dimmer than their nuclear QSO light
(in apparent magnitudes in the observed wavebands).  Very few of the
objects have hosts nearly this much dimmer than their nuclei, and
there are in fact only 7 objects that have a host 2.1 or more
magnitudes dimmer than their
nucleus.  In general, considering that the
host magnitudes span a noticeably smaller range than the nuclear
magnitudes (Figure~\ref{fig:MvsM}), combined with the fact that we
have failed to detect a host in only one case, makes us confident that
our host luminosity function is not strongly biased by missing very
dim hosts.

%and we find that this holds true for host sizes ranging from smaller
%than $ \approx 0\farcs5 $ to $ \approx 4\farcs0 $, beyond which there
%is little relevant data.

%  Considering the faint end of the host distribution in
%particular, the commonly accepted division between QSOs and less
%luminous AGN is $M_V \leq -23$ (for $H_0=50$ km s$^{-1}$ Mpc$^{-1}$), 
%and our data
%suggest that we could see a host as dim as $M_V \approx -19$ to $ -20
%$ for a QSO of this magnitude. In general, considering that the host
%magnitudes span a noticeably smaller range than the nuclear QSO
%magnitudes (Figure 1), combined with the fact that we have failed to
%detect a host in only two cases so far, makes us confident that our
%host luminosity function is not strongly biased by missing very dim
%hosts.

The ellipticals outnumber the spirals in our sample, making up 62\% of
the total.  However, due to {\it HST} target selection effects, this
may not be representative of all QSOs in this redshift range.  For
example, we confirm, with only one or two possible exceptions, that
radio-loud QSOs are found in elliptical or interacting hosts
(\S\ref{sec:unweighted}), while radio-quiet QSOs may be found in
either ellipticals, spirals or interacting cases.  The fraction of
radio-loud QSOs in our sample is $\approx 37\%$, higher than the
10--20\% expected for optically--selected samples in this redshift
range (Kellermann et al.~1989; Hooper et al.~1995; Hooper et
al.~1996).  This may indicate that we have selected an artificially
high fraction of QSOs with elliptical hosts.  But since QSOs in
ellipticals are more luminous than those in spirals, our weighting
procedure (\S\ref{sec:weightdist}) may compensate for this bias to an
extent.

It is unlikely that a redshift-dependent magnitude bias could arise
from non-stellar emission from the hosts.  Spiral hosts may contain H
{\sc ii} regions, however most bright H {\sc ii} regions are masked
from our fits and analyses if they are in dim areas of the
host. Although less-prominent H {\sc ii} regions might be unmasked,
the rest equivalent width of H$\alpha$, the major H {\sc ii} emission
line, is typically $\sim 25\,$ \AA \, in late-type galaxies (Gavazzi
et al.~1998).  In contrast, the WFPC2 wide-band filters we use have
much larger equivalent widths ranging from 867 \AA \, (F675W) to 1539
\AA \, (F814W) (Burrows~1995), so the H$\alpha$ effect in the host is
very small. More important is the affect that broad emission lines
might have on a QSO's nuclear luminosity, and H$\alpha$ is again the
most prominent line.  In this case, H$\alpha$ rest-frame equivalent
widths are distributed with a median of $\approx 240$ \AA \, and
usually do not exceed $\approx 450$ \AA \, (Sabbey~1999); thus this
could account for at most a few tenths of a magnitude variation,
depending on whether or not the line is included in the filter
passband.

\subsection{Comparisons with Results from Some Other 
Studies}\label{sec:comparisons}

There have been a few other large-sample studies of QSO hosts with
selection criteria similar to our own.  Of the space-based {\it HST}
ones, the two largest, those of Bahcall et al.~(1997) and of McLure et
al.~(1999), have samples that are included in ours.

The sample of Bahcall et al.~(1997) includes 20 QSOs with redshifts $z
<0.3$ and $M_V < -24.4$, making them among the most luminous objects
in the nearby universe.  In this comparison, we use their final
results, which come from their 2-D model fits.  Their host magnitude
distribution has a shape similar to ours but is nearly a magnitude
fainter.  It should be noted that the results of their 1-D fits are
brighter than their 2-D fits and give a distribution about half a
magnitude fainter than ours.  They find that, on average, the hosts of
radio-loud QSOs are one magnitude brighter than the hosts of
radio-quiet QSOs, while we find that the radio-loud hosts are about
half a magnitude brighter.  They also report the hint of a luminosity
difference between elliptical and spiral hosts, but they state that it
may be artificial, a consequence of fitting the host model to the
outer ($r\geq 1\farcs0$) region of the host.  Our analysis shows only
a small difference, with elliptical hosts being $\approx 0.2$ mag
brighter than the spirals on average.  This is true whether using the
mean or the median.

Bahcall et al.~(1997) classify 4 of their hosts as spirals, 12 as
ellipticals, and the remaining 4 as interacting or of indeterminate
morphology.  Ignoring the interacting and indeterminate types, we
agree with their morphology classifications, with the possible
exceptions of PG~1116$+$215 and PG~1444$+$407.  Bahcall et al.~(1997)
classify both of these as elliptical, but we find they are disk-like.
In our analysis, they are best fit by exponential profiles, and they
contain central bulges, although neither host shows visual evidence of
spiral arms.  One of the central conclusions of Bahcall et al.~(1997),
that elliptical galaxies can host either radio-loud or radio-quiet
QSOs, is confirmed by our study.  Finally, we agree with their finding
that the host magnitude distribution is inconsistent with a Schechter
(1976) function.

The sample of McLure et al.~(1999) includes 15 QSOs with a redshift
range of $0.1 \leq z \leq 0.35$.  Transforming their results into
$M_V$, using $\vr=0.7$ for ellipticals and $\vr=0.6$ for spirals
(Fukugita et al.~1995), we find that their host luminosity
distribution is narrower ($ -21.9 > M_V > -23.6$) than ours, though
with the peak in the same magnitude bin ($M_V \sim -23$) as ours.
They find only two QSOs with spiral hosts (with the remaining 13 being
ellipticals), while we have a much larger fraction of spirals, 26/70.
But note that we classify MRK~1014 as a spiral on the basis of its
arms, while they list it as an elliptical on the basis of its
$r^{1/4}$-law radial profile.  While McLure et al.~(1999) find that
essentially all radio-quiet QSOs with nuclear luminosities $M_V <
-23.7$ ($M_R < -24.0$) have elliptical hosts, we find several spiral
hosts with nuclei in this luminosity range.  However, our data do show
that the radio-quiet QSOs with the brightest nuclei reside in
elliptical hosts, while those with the dimmest nuclei reside in spiral
hosts.  This may still lend support to their idea that the correlation
between black hole and bulge mass derived by Magorrian et al.~(1998)
affects the distribution of nuclear luminosities between elliptical
and spiral hosts.

Since this investigation uses archival {\it HST} images, we also
compare our host apparent filter magnitudes, $m_{\mathrm{host}}$,
against those of the above two studies, for cases in which we use the
same images.  This allows a comparison of the measurement techniques
themselves.  For the Bahcall data, this corresponds to 16 objects from
{\it HST} observing programs numbered 5099, 5343, and 5849 (see
Table~\ref{table:obslist} for the object names).  For the McLure data,
this corresponds to 11 objects from program number 6776 (but note that
PKS 2135$-$147 is not included in their paper).  We note in particular
that Bahcall et al.~(1997) provide the results of both their
one-dimensional and two-dimensional model fits; while they adopt the
two-dimensional magnitudes in their analysis, we find that their
one-dimensional magnitudes agree better with ours, and thus adopt
their one-dimensional magnitudes in the following comparison.  There
are 27 other objects for which there exist published
$m_{\mathrm{host}}$ values taken from the same observations we use.
These include the 9 objects observed by Boyle (program 6361) and
published in Schade et al.~(2000), the 10 observed by Impey (program
5450) and published in Hooper, Impey, \& Foltz~(1997), the four
observed by Disney (program 6303) and published in Boyce et al.~(1998)
[note that we exclude PG~0043$+$039, for which the published host
measurements are very uncertain, and IR~0450$-$2958 and
IR~0759$+$6508, which are published in Boyce et al.~(1996) without
values for $m_{\mathrm{host}}$], the two observed by Hutchings
(program 5178) and published in Hutchings \& Morris~(1995), and the
two observed by Macchetto (program 5143) and published in Disney et
al.~(1995) and Boyce et al.~(1998).

The comparisons show that the one-dimensional $m_{\mathrm{host}}$ data
of Bahcall et al.~(1997) are on average $0.09 \pm 0.35$ mag fainter
than ours, and the $m_{\mathrm{host}}$ data of McLure et al.~(1999)
are an average of $0.16 \pm 0.33$ mag fainter than ours.  Looking at
all 54 objects together, we find that published $m_{\mathrm{host}}$
results average $0.30 \pm 0.62$ magnitudes fainter than ours.
Although there is considerable scatter to these differences in
apparent filter magnitudes, we believe that at least part of the
difference may be systematic and lie in our direct measurement of the
apparent host magnitude from the PSF-subtracted image, without relying
on a simple galaxy model except at large radii.  In theory, if the
host model fit is weighted inversely to the square of the Poisson
noise in the image ($1/\sigma$), the model's magnitude will be
slightly fainter than that of the actual host image.  In practice, we
find that our host models are $\approx 0.25$ mag fainter than the host
magnitudes we calculate in \S\ref{sec:extractmag}.  The host
magnitudes we adopt (\S\ref{sec:extractmag}) are not affected much by
this bias in the models.

\section {CONCLUSIONS}\label{sec:conclusions}

1. We have assembled a sample of 71 {\it HST} WFPC2 imaging
observations of luminous QSOs (total nuclear plus host light $M_V \leq
-23$ in our adopted cosmology with $H_0 = 50$ km s$^{-1}$ Mpc$^{-1}$,
$q_0 = 0.5$, and $\Lambda = 0$) in the redshift interval $0.06 \leq z
\leq 0.46$.  We derive results on QSO host and nuclear luminosities
and on host morphology, using procedures we have developed, and we
compile results on radio loudness.  Of the 71 QSOs, we detect hosts in
70 cases.  The one non-detection may be due to filter choice (F336W)
and faintness.

2. The host galaxies span a narrow range of luminosities and are
exceptionally bright, much more so than normal galaxies, usually $L >
L^*_V$.

3. The hosts are almost equally divided between subclasses of
radio-loud QSOs with elliptical hosts (22 objects), radio-quiet QSOs
with elliptical hosts (22 objects), and radio-quiet QSOs with spiral
hosts (21 objects).  Radio-loud QSOs with spiral hosts (at most 4
objects) are extremely rare.  

4. The elliptical host luminosity distribution of the radio-loud QSOs
differs significantly from both the elliptical and spiral host
luminosity distributions of the radio-quiet QSOs. However, the latter
two distributions are more compatible.  Spiral hosts are typically
nearly as luminous as elliptical hosts, and the hosts of radio-loud
QSOs are typically 0.5 magnitude brighter than those of radio-quiet
QSOs.

5. Using a weighting procedure, we derive the combined luminosity
function of low-redshift QSO host galaxies. Subject to systematic
uncertainties in normalization procedures, the luminosity function of
nearby QSO hosts peaks near the point where the normal galaxy
luminosity function falls off.  We conclude that the host luminosity
function of low-redshift QSOs differs in shape from the normal galaxy
luminosity function but that they coincide at the highest
luminosities.  With a QSO defined in historically traditional terms,
i.e., total nuclear plus host light has $M_V \leq -23$, the ratio of
the number of nearby QSO hosts to the number of normal galaxies is
$\mathcal{R} = (L_V/{11.48L^*_V})^{2.46}$, where $L_V^*$ corresponds
to $M_V^*=-22.35$.  This ratio represents an empirical
parameterization of the probability that a galaxy with luminosity
$L_V$ will host a QSO at redshift $z\approx0.26$.

\acknowledgements

This research has been supported by a Graduate Student Fellowship and
a grant from the Director's Discretionary Research Fund, both from the
Space Telescope Science Institute.  We made use of the following
databases: the HST data archive; NASA/IPAC Extragalactic Database
(NED), operated by the Jet Propulsion Laboratory, California Institute
of Technology, under contract with the National Aeronautics and Space
Administration; the SIMBAD database, operated at CDS, Strasbourg,
France; and NASA's Astrophysics Data System Abstract Service.  We
would like to thank Sandhya Rao for discussions about luminosity
functions and comments on the manuscript, as well as Chris O'Dea
for his helpful suggestions.

\clearpage

%%%%%%%%%%%%%%%%%%%%%%%%%%%%%%%%%%%%%%%%%%%%%%%%%%%%%%%%%%%%%%%%%%%%%%%%%%%%%%%%%%%%%%%%%%%%%%%%%%%%%%%%%%%%%%%%%%%%%%%%%%%%%%%%%%%%%%%%

%%%jeeves~/papers/paper1/submission3/data/four-prof-diff-paper.ps

%%%figs/lf/unweighted/mnuc-mhost1.ps
%%%old f1.eps

%%% Replaces part 2 of fig:unwLF
%%% unwhostdist_paper_v1_sub.ps
%%% old f3.eps

%%% Replaces part 1 of fig:unwLF.
%%% unwhostdist_paper_v1_all.ps
%%% old f2.eps

%%% Replaces above fig:wLF
%%% lf_mhost_weighted_wext_ropt_paper.ps
%%% old f4.eps

%%% figs/lf/normalized/lf_norm.ps
%%% previously f4.eps
%%% old f5.eps

%%% figs/lf/normalized/lf_ratio.ps
%%% previously f5.eps
%%% old f6.eps

%%% figs/qsoplots/3c93-prof.ps
%%% 3c93-prof-paper.ps
%%% old f7.eps

%%% figs/qsoplots/3c48-prof.ps
%%% 3c48-prof-paper.ps
%%% old f8.eps

%%% pks2349-prof-paper.ps
%%% old f9.eps

%%% ms1059-prof-paper.ps
%%% old f10.eps

%%%%%%%%%%%%%%%%%%%%%%%%%%%%%%%%%%%%%%%%%%%%%%%%%%%%%%%%%%%

\begin{deluxetable}{llllccrccccccl}
 %\rotate
 \tabletypesize {\tiny}
 %\tabletypesize {\scriptsize}
 %\setlength{\tabcolsep}{0.04in}
 \setlength{\tabcolsep}{0.05in}
 %\setlength{\tabcolsep}{0.01in}
 %\tablefontsize{\tiny}
 %\small
 %\scriptsize
 %\footnotesize <--good one
 %\tiny
 %\normalsize
 \tablewidth{0pt}
 %Note: \renewcommand{\arraystretch}{x} puts a space between table rows that
 %      is x times the normal spacing:
 \renewcommand{\arraystretch}{.6}

\tablecaption{Observations and Data}

\tablehead{
\colhead{$\begin{array}[t]{c} \mbox{RA} \\ \mbox{(J2000)} \\ (1) \end{array}$}&
\colhead{$\begin{array}[t]{c} \mbox{Dec} \\ \mbox{(J2000)} \\ (2) \end{array}$}&
\colhead{$\begin{array}[t]{c} \mbox{Name} \\ \\ (3) \end{array}$}&
\colhead{$\begin{array}[t]{c} \!\!\!\!\! z \!\!\!\!\! \\ \!\!\!\!\!\!\!\!\!\! \\ \!\!\!\!\! (4) \!\!\!\!\! \end{array}$}&
\colhead{$\begin{array}[t]{c} \!\!\!\!\! \mbox{CCD} \!\!\!\!\! \\ \!\!\!\!\!\!\!\!\!\! \\ \!\!\!\!\! (5) \!\!\!\!\! \end{array}$}&
\colhead{$\begin{array}[t]{c} \!\!\!\!\! \mbox{Filter} \!\!\!\!\! \\ \!\!\!\!\!\!\!\!\!\! \\ \!\!\!\!\! (6) \!\!\!\!\! \end{array}$}&
\colhead{$\begin{array}[t]{c} \!\!\!\!\! t_{\mathrm{exp}} \!\!\!\!\! \\ \!\!\!\!\! \mbox{(sec)} \!\!\!\!\! \\ \!\!\!\!\! (7) \!\!\!\!\! \end{array}$}&
\colhead{$\begin{array}[t]{c} \!\!\!\!\! m_{\mathrm{nuc}} \!\!\!\!\! \\ \!\!\!\!\!\!\!\!\!\! \\ \!\!\!\!\! (8) \!\!\!\!\! \end{array}$}&
\colhead{$\begin{array}[t]{c} \!\!\!\!\! m_{\mathrm{host}} \!\!\!\!\! \\ \!\!\!\!\!\!\!\!\!\! \\ \!\!\!\!\! (9) \!\!\!\!\! \end{array}$}&
\colhead{$\begin{array}[t]{c} \!\!\!\!\! M_V\mathrm{(nuc)} \!\!\!\!\! \\ \!\!\!\!\!\!\!\!\!\! \\ \!\!\!\!\! (10) \!\!\!\!\! \end{array}$}&
\colhead{$\begin{array}[t]{c} \!\!\!\!\! M_V\mathrm{(host)} \!\!\!\!\! \\ \!\!\!\!\!\!\!\!\!\! \\ \!\!\!\!\! (11) \!\!\!\!\! \end{array}$}&
\colhead{$\begin{array}[t]{c} \!\!\!\!\! \mbox{Mph.} \!\!\!\!\! \\ \!\!\!\!\!\!\!\!\!\! \\ \!\!\!\!\! (12) \!\!\!\!\! \end{array}$}&
\colhead{$\begin{array}[t]{c} \!\!\!\!\! \mbox{Rad.} \!\!\!\!\! \\ \!\!\!\!\!\!\!\!\!\! \\ \!\!\!\!\! (13) \!\!\!\!\! \end{array}$}&
\colhead{$\begin{array}[t]{c} \!\!\!\!\! \mbox{Prog.}\!\!\!\!\! \\ \!\!\!\!\!\!\!\!\!\! \\ \!\!\!\!\! (14) \!\!\!\!\! \end{array}$}
}

\startdata

00 23 11.1&$+$00 35 16.5  &LBQS 0020$+$0018  &0.423   &PC1 &F675W  &1200  &19.30  &19.34  &-22.45   &-22.77  &E    &Q   &5450 \\
00 24 03.7&$-$02 45 27.8  &LBQS 0021$-$0301  &0.422   &PC1 &F675W  &1200  &19.03  &19.10  &-22.74   &-23.03  &E    &Q   &5450 \\
00 24 32.5&$-$29 28 55.5  &MRC 0022$-$297 \tablenotemark{a}     &0.406   &PC1 &F336W  &2100  &...    &...    & ...     & ...    &...  &L   &5974 \\
00 45 46.6&$+$04 11 15.8  &PG 0043$+$039      &0.385   &PC1 &F702W  &1800  &16.04  &19.03  &-25.49   &-22.67  &E    &Q   &6303 \\
00 54 55.7&$+$25 25 47.2  &PG 0052$+$251      &0.155   &WF3 &F606W  &2100  &16.04  &16.83  &-23.80   &-23.08  &S    &Q   &5343 \\
00 57 11.6&$+$14 45 24.5  &PHL 909            &0.171   &WF3 &F606W  &2100  &15.97  &16.89  &-24.07   &-23.29  &E    &Q   &5343 \\
01 03 13.0&$+$02 21 10.5  &UM 301             &0.393   &PC1 &F675W  &1280  &17.66  &19.44  &-23.92   &-22.44  &E    &Q   &5450 \\
01 36 22.4&$+$20 57 14.7  &3C 47              &0.425   &PC1 &F702W  &280   &17.82  &18.75  &-24.07   &-23.37  &E    &L   &5476 \\
01 37 39.0&$+$33 09 22.0  &3C 48              &0.367   &PC1 &F814W  &3500  &15.74  &16.18  &-25.72   &-24.83  &EI   &L   &5235 \\
01 39 59.0&$+$01 31 01.6  &PHL 1093           &0.26    &WF2 &F675W  &1871  &17.21  &17.17  &-23.47   &-23.54  &E    &L   &6776 \\
01 59 52.2&$+$00 23 50.1  &MRK 1014           &0.163   &PC1 &F814W  &480   &16.17  &14.75  &-23.30   &-24.34  &S    &Q   &5982 \\
02 02 05.1&$-$76 20 04.3  &PKS 0202$-$76      &0.389   &PC1 &F702W  &1800  &16.67  &18.72  &-24.98   &-23.12  &E    &L   &6303 \\
02 07 49.5&$+$02 43 43.0  &NAB 0205$+$02      &0.155   &WF3 &F606W  &2100  &15.40  &18.08  &-24.37   &-21.77  &S    &Q   &5343 \\
02 47 42.2&$+$19 40 09.4  &Q 0244$+$194       &0.176   &WF2 &F675W  &1871  &16.80  &17.54  &-23.29   &-22.47  &E    &Q   &6776 \\
03 00 31.6&$+$02 40 06.7  &US 3498            &0.115   &WF2 &F675W  &1871  &19.30  &15.87  &-19.79   &-23.10  &S    &Q   &6776 \\
03 11 46.7&$-$76 51 40.8  &PKS 0312$-$77      &0.223   &PC1 &F702W  &1800  &16.13  &16.67  &-24.40   &-23.78  &E    &L   &6303 \\
03 18 07.9&$-$34 25 52.2  &Q 0316$-$346       &0.260   &WF3 &F606W  &2100  &16.21  &18.08  &-24.68   &-23.03  &IS   &Q   &5343 \\
03 43 30.0&$+$04 57 48.6  &3C 93              &0.357   &PC1 &F702W  &280   &18.58  &18.49  &-23.51   &-23.73  &E    &L   &5476 \\
04 52 32.4&$-$29 53 41.0  &IR 0450$-$2958     &0.286   &PC1 &F702W  &1800  &15.40  &17.17  &-25.41   &-23.65  &SI   &Q   &6303 \\
07 39 18.0&$+$01 37 04.6  &PKS 0736$+$01      &0.191   &WF2 &F675W  &1871  &16.30  &16.70  &-24.03   &-23.58  &E    &L   &6776 \\
07 57 57.8&$+$39 20 34.7  &MS 07546$+$3928    &0.096   &PC1 &F814W  &610   &14.26  &14.37  &-24.22   &-23.47  &E    &Q   &6361 \\
08 04 35.3&$+$64 59 53.9  &IR 0759$+$6508     &0.149   &PC1 &F702W  &1800  &15.94  &15.65  &-23.57   &-23.73  &SI   &Q   &6303 \\
08 04 55.0&$+$21 20 45.7  &MS 0801.9$+$2129   &0.118   &PC1 &F814W  &610   &16.00  &15.66  &-22.91   &-22.80  &S    &Q   &6361 \\
08 39 52.6&$-$12 14 42.7  &3C 206             &0.198   &PC1 &F702W  &600   &16.07  &16.90  &-24.03   &-23.09  &E    &L   &5957 \\
09 06 31.9&$+$16 46 11.5  &3C 215             &0.412   &PC1 &F814W  &5000  &17.71  &18.23  &-24.08   &-23.08  &E    &L   &5988 \\
09 25 57.7&$+$19 53 45.4  &PG 0923$+$201      &0.19    &WF3 &F606W  &2100  &15.53  &17.46  &-24.73   &-23.00  &E    &Q   &5343 \\
09 46 50.7&$+$13 19 52.6  &MS 0944.1$+$1333   &0.131   &PC1 &F814W  &600   &14.89  &15.93  &-24.16   &-22.52  &E    &Q   &6361 \\
09 56 48.7&$+$41 15 47.2  &PG 0953$+$414      &0.234   &WF2 &F675W  &1991  &15.17  &17.21  &-25.23   &-23.22  &S    &Q   &6776 \\
10 04 00.4&$+$28 55 20.2  &PG 1001$+$291      &0.330   &WF3 &F702W  &2400  &15.59  &17.90  &-25.58   &-23.33  &S    &Q   &5949 \\
10 07 29.1&$+$12 48 33.3  &PKS 1004$+$13      &0.24    &WF3 &F606W  &2100  &15.15  &17.00  &-25.62   &-24.06  &E    &L   &5343 \\
10 14 56.2&$+$00 34 21.2  &PG 1012$+$008      &0.185   &WF2 &F675W  &1931  &16.22  &16.76  &-23.75   &-23.19  &SI   &Q   &6776 \\
10 31 52.5&$-$14 16 10.9  &HE 1029$-$1401     &0.086   &WF3 &F606W  &2100  &13.84  &15.86  &-24.79   &-22.74  &E    &Q   &5343 \\
11 02 38.2&$+$72 46 09.9  &MS 1059.0$+$7302   &0.089   &PC1 &F814W  &600   &16.60  &15.41  &-21.65   &-22.34  &S    &Q   &6361 \\
11 19 06.7&$+$21 18 39.3  &PG 1116$+$215      &0.177   &WF3 &F606W  &1800  &14.85  &16.74  &-25.19   &-23.42  &S    &Q   &5099 \\
12 04 42.2&$+$27 54 12.0  &PG 1202$+$281      &0.165   &WF3 &F606W  &1800  &16.85  &17.39  &-23.03   &-22.62  &E    &Q   &5099 \\
12 12 27.9&$+$12 42 54.5  &LBQS 1209$+$1259   &0.418   &PC1 &F675W  &1200  &19.35  &19.38  &-22.39   &-22.71  &E    &Q   &5450 \\
12 19 23.1&$+$06 38 26.8  &PG 1216$+$069      &0.331   &PC1 &F702W  &1800  &15.42  &18.70  &-25.76   &-22.55  &E    &Q   &5143 \\
12 20 37.2&$+$17 18 24.4  &LBQS 1218$+$1734   &0.444   &PC1 &F675W  &1200  &18.33  &19.01  &-23.54   &-23.26  &E    &L   &5450 \\
12 21 45.9&$+$75 19 06.5  &MS 1219.6$+$7535   &0.071   &PC1 &F814W  &610   &15.06  &14.56  &-22.72   &-22.52  &ED   &Q   &6361 \\
12 25 10.7&$+$09 54 38.8  &LBQS 1222$+$1010   &0.398   &PC1 &F675W  &1200  &18.38  &18.62  &-23.23   &-23.24  &S    &Q   &5450 \\
12 25 15.0&$+$12 18 40.2  &LBQS 1222$+$1235   &0.412   &PC1 &F675W  &1200  &17.68  &18.25  &-24.04   &-23.82  &E    &L   &5450 \\
12 29 09.9&$+$02 03 02.3  &3C 273             &0.158   &WF3 &F606W  &1800  &12.60  &15.65  &-27.19   &-24.24  &E    &L   &5099 \\
12 32 03.6&$+$20 09 29.2  &PG 1229$+$204      &0.064   &PC1 &F702W  &560   &15.37  &15.04  &-22.27   &-22.33  &S    &Q   &5502 \\
12 42 39.5&$+$17 38 22.6  &LBQS 1240$+$1754   &0.458   &PC1 &F675W  &1480  &17.98  &19.31  &-23.93   &-23.02  &E    &Q   &5450 \\
12 46 30.2&$+$16 45 23.5  &LBQS 1243$+$1701   &0.459   &PC1 &F675W  &1400  &18.45  &18.44  &-23.49   &-23.91  &E    &Q   &5450 \\
12 52 25.2&$+$56 34 36.4  &3C 277.1           &0.321   &PC1 &F702W  &280   &17.97  &18.35  &-23.09   &-22.76  &S    &L   &5476 \\
13 05 36.1&$-$10 33 36.2  &PG 1302$-$102      &0.278   &WF3 &F606W  &1800  &15.19  &17.35  &-25.93   &-24.14  &E    &L   &5099 \\
13 09 47.0&$+$08 19 49.5  &PG 1307$+$085      &0.155   &WF3 &F606W  &1800  &15.46  &17.47  &-24.33   &-22.42  &E    &Q   &5343 \\
13 12 16.3&$+$35 14 36.7  &PG 1309$+$355      &0.184   &WF3 &F606W  &2100  &15.56  &16.61  &-24.53   &-23.62  &S    &L   &5343 \\ %\tablebreak
14 00 33.9&$+$04 04 46.8  &PG 1358$+$04       &0.427   &PC1 &F702W  &1800  &15.96  &18.02  &-25.84   &-24.02  &E    &Q   &6303 \\
14 04 38.7&$+$43 27 07.5  &Q 1402$+$436       &0.323   &PC1 &F702W  &560   &15.15  &17.42  &-25.93   &-23.73  &EI   &Q   &5178 \\
14 05 12.9&$+$25 55 17.7  &PG 1402$+$261      &0.164   &WF3 &F606W  &2100  &15.73  &17.33  &-24.12   &-22.62  &S    &Q   &5343 \\
14 19 05.7&$-$13 10 56.5  &MS 1416.3$-$1257   &0.129   &PC1 &F814W  &600   &15.83  &16.90  &-23.37   &-21.70  &E    &Q   &6361 \\
14 27 33.6&$+$26 32 52.9  &B2 1425$+$267      &0.366   &WF3 &F814W  &3500  &15.88  &17.47  &-25.49   &-23.45  &E    &L   &5235 \\
14 29 08.6&$+$01 17 13.0  &MS 1426.5$+$0130   &0.086   &PC1 &F814W  &610   &14.30  &14.49  &-23.87   &-23.17  &S    &Q   &6361 \\
14 46 49.1&$+$40 34 34.7  &PG 1444$+$407      &0.267   &WF3 &F606W  &2326  &15.80  &17.37  &-25.14   &-23.81  &S    &Q   &5849 \\
15 14 39.2&$+$36 50 37.7  &B2 1512$+$37       &0.371   &WF3 &F814W  &1600  &16.04  &17.31  &-25.38   &-23.66  &E    &L   &6490 \\
15 22 30.7&$-$06 44 43.1  &MS 1519.8$-$0633   &0.083   &PC1 &F814W  &600   &16.01  &15.07  &-22.32   &-22.75  &S    &Q   &6361 \\
15 47 47.5&$+$20 51 33.1  &3C 323.1           &0.264   &WF3 &F606W  &1800  &16.07  &18.01  &-24.94   &-23.33  &E    &L   &5099 \\
15 50 42.5&$+$11 19 54.2  &MC 1548$+$114A     &0.436   &WF3 &F702W  &1400  &18.27  &19.92  &-23.66   &-22.20  &SI   &L   &5682 \\
16 37 46.5&$+$11 49 49.7  &MC 1635$+$119      &0.146   &WF2 &F675W  &1931  &18.12  &16.73  &-21.38   &-22.62  &E    &Q   &6776 \\
17 04 38.3&$+$60 44 51.4  &3C 351             &0.372   &PC1 &F702W  &1800  &15.50  &16.97  &-25.96   &-24.59  &S    &L   &6303 \\
21 37 48.1&$-$14 32 30.9  &PKS 2135$-$147     &0.200   &WF2 &F675W  &1931  &16.21  &16.91  &-23.96   &-23.23  &E    &L   &6776 \\
21 43 38.3&$+$17 43 14.2  &OX 169             &0.211   &WF2 &F675W  &1871  &15.89  &17.28  &-24.59   &-23.18  &EI   &L   &6776 \\
22 02 56.6&$-$56 59 10.7  &MS 2159.5$-$5713   &0.083   &PC1 &F814W  &610   &17.14  &15.01  &-20.91   &-22.52  &S    &?   &6361 \\
22 03 15.0&$+$31 45 38.3  &Q 2201$+$315       &0.295   &PC1 &F702W  &560   &15.46  &16.75  &-25.78   &-24.50  &E    &L   &5178 \\
22 16 51.7&$-$18 48 14.0  &LBQS 2214$-$1903   &0.396   &PC1 &F675W  &1280  &18.81  &19.27  &-22.81   &-22.60  &S    &Q   &5450 \\
22 17 45.8&$-$03 32 47.1  &Q 2215$-$037       &0.242   &PC1 &F702W  &1800  &18.69  &17.38  &-22.06   &-23.29  &E    &Q   &5143 \\
22 50 27.5&$+$14 19 09.7  &PKS 2247$+$14      &0.237   &WF2 &F675W  &1871  &16.65  &17.22  &-23.90   &-23.32  &E    &L   &6776 \\
23 47 27.6&$+$18 44 06.9  &Q 2344$+$184       &0.138   &WF2 &F675W  &1871  &20.22  &16.68  &-19.16   &-22.60  &S    &Q   &6776 \\
23 51 53.0&$-$01 09 27.8  &PKS 2349$-$014     &0.174   &WF2 &F675W  &1871  &15.97  &15.63  &-23.82   &-24.07  &IE   &L   &6776 \\

\enddata
\label{table:obslist} 

\tablecomments {Col. (1), RA in {\it hh mm ss.s}.  Col. (2), Dec in
{\it dd mm ss.s}.  Col. (7), Exposure time.  Col. (8), apparent
nuclear magnitude in filter.  Col. (9), apparent host magnitude in
filter.  Col. (10), absolute $V$ nuclear magnitude.  Col. (11),
absolute $V$ host magnitude.  Col. (12), host morphology: a)
E=elliptical; b) S=spiral; c) EI=elliptical undergoing strong
interaction; d) SI=spiral undergoing strong interaction; e)
ED=elliptical with possible inner disk; f) IE=irregular or interacting
that is best fit with an elliptical model; g) IS=irregular or
interacting that is best fit with a spiral model.  Col. (13),
radio-loudness: Q = radio-quiet; L = radio-loud; ? = radio-loudness
not available.  Col. (14), observing programs and principal
investigators: a) 5099=Bahcall; b) 5143=Macchetto; c) 5178=Hutchings;
d) 5235=Westphal; e) 5343=Bahcall; f) 5450=Impey; g) 5476=Sparks; h)
5502=Sparks; i) 5682=Burbidge; j) 5849=Bahcall; k) 5949=Lanzetta; l)
5957=Sparks; m) 5974=Lehnert; n) 5982=Sanders; o) 5988=Ellingson; p)
6303=Disney; q) 6361=Boyle; r) 6490=Stockton; s) 6776=Dunlop.}

\tablenotetext {a} {Host not detected.}

\end{deluxetable}
%%%%%%%%%%%%%%%%%%%%%%%%%%%%%%%%%%%%%%%%%%%%%%%%%%%%%%%%%%%%%%%%%%%%%%%%%%%%%%%%%%%%%%%%%%%%%%%%%%%%%%%%%%%%%%%%%%%%%%%%%%%%%%%%%%%%%%%

\begin{deluxetable}{llll}
\footnotesize
\tablewidth{0pt}
\renewcommand{\arraystretch}{.8}

\tablecaption{Median Absolute Magnitudes of Subclasses}

\tablehead{
\colhead{Subclasses \tablenotemark{a}}&
\colhead{No. members}&
\colhead{$\begin{array}[t]{c} \mbox{Median} \\ M_V(\mathrm{host}) \end{array}$}&
\colhead{$\begin{array}[t]{c} \mbox{Median} \\ M_V(\mathrm{nuc}) \end{array}$}
}

\startdata

All  & 70 & -23.18  & -24.03 \\
LE   & 22 & -23.54  & -24.08 \\
QE   & 22 & -22.71  & -23.92 \\
LS   &  4 & -22.76  & -23.66 \\
QS   & 21 & -23.10  & -23.75 \\
L    & 26 & -23.54  & -24.08 \\
Q    & 43 & -23.00  & -23.80 \\
E    & 44 & -23.26  & -24.07 \\
S    & 26 & -23.08  & -23.30 \\

\enddata
\label{table:medians}

\tablenotetext {a} {LE=radio-loud QSOs in elliptical hosts;
QE=radio-quiet QSOs in elliptical hosts; 
QS=radio-quiet QSOs in spiral hosts; 
L=radio-loud QSOs;
Q=radio-quiet QSOs; 
E=elliptical hosts; 
S=spiral hosts.}

\end{deluxetable}

%%%%%%%%%%%%%%%%%%%%%%%%%%%%%%%%%%%%%%%%%%%%%%%%%%%%%%%%%%%%%%%%%%%%%%%%%%%%%%%%%%%%%%%%%%%%%%%%%%%%%%%%%%%%%%%%%%%%%%%%%%%%%%%%%%%%%%%
\begin{deluxetable}{lllll}
\footnotesize
\tablewidth{0pt}
\renewcommand{\arraystretch}{.8}

\tablecaption{Kolmogorov-Smirnov Test Results for Subclass Comparison}

\tablehead{
\colhead{Subclasses (number)\tablenotemark{a}}&
\colhead{$D_{\mathrm{host}}$}&
\colhead{$p_{\mathrm{host}}$}&
\colhead{$D_{\mathrm{nuc}}$}&
\colhead{$p_{\mathrm{nuc}}$}
}
\startdata

LE (22), QE (22)      & 0.682 & $2.75\times10^{-5}$ & 0.409 & 0.0356 \\
LE (22), QS (21)      & 0.487 & $7.50\times10^{-3}$ & 0.437 & 0.0222 \\
QE (22), QS (21)      & 0.346 & $1.19\times10^{-1}$ & 0.165 & 0.908  \\
L (26), Q (43)        & 0.504 & $2.83\times10^{-4}$ & 0.380 & 0.0131 \\
E (44), S (26)        & 0.215 & $3.91\times10^{-1}$ & 0.297 & 0.0900 \\

\enddata
\label{table:kstests}

\tablenotetext {a} {LE=radio-loud QSOs in elliptical hosts;
QE=radio-quiet QSOs in elliptical hosts; QS=radio-quiet QSOs in spiral hosts; 
L=radio-loud QSOs;
Q=radio-quiet QSOs; 
E=elliptical hosts; 
S=spiral hosts.}

\tablecomments {The parameter $D$ is the K-S statistic. The
parameter $p$ is the probability of obtaining $D$ if the
objects in both subclasses are drawn from the same parent 
population.}

\end{deluxetable}


\clearpage

\begin{thebibliography}{}

\bibitem[Bahcall et al.(1997)]{Bahcall97} Bahcall, J.  N., Kirhakos, S., \& Saxe, D.  H. 1997, \apj, 479, 642

\bibitem[Blanton et al.(2001)]{Blanton01} Blanton, M. R., \& SDSS Collaboration.  2001, preprint (astro-ph/0012085)

\bibitem[Boroson et al.(1985)]{Boroson85} Boroson, T. A., Persson, S. E., \& Oke, J. B. 1985, \apj, 293, 120

\bibitem[Boyce et al.(1996)]{Boyce96} Boyce, P., et al. 1996, \apj, 473, 760

\bibitem[Boyce et al.(1998)]{Boyce98} Boyce, P. J., et al. 1998, \mnras, 298, 121

% \bibitem[Boyce et al.(1999)]{Boyce99} Boyce, P., Disney, M., \& Bleaken, D. 1999, \mnras, 302, L39

\bibitem[Boyle et al.(2000)]{Boyle00} Boyle, B. J., Shanks, T., Croom,
S. M., Smith, R. J., Miller, L., Loaring, N., \& Heymans, C. 2000,
\mnras, 317, 1014

\bibitem[Brinkmann et al.(1997)]{Brinkmann97} Brinkmann, W., Yuan, W.,
\& Siebert, J. 1997, \aap, 319, 413

\bibitem[Burrows(1995)]{Burrows95} Burrows, C. J., ed. 1995, Wide
Field and Planetary Camera 2 Instrument Handbook (Baltimore: STScI)

\bibitem[Cristiani \& Vio(1990)]{CV90} Cristiani, S. \& Vio, R. 1990,
\aap, 227, 385

\bibitem[De Vries et al.(1999)]{DeVries99} De Vries, W. H., O'Dea,
C. P., Baum, S. A., \& Barthel, P. D. 1999, \apj, 526, 27

\bibitem[Disney et al.(1995)]{Disney95} Disney, M. J., et al. 1995,
Nature, 376, 150

\bibitem[Dunlop et al.(1993)]{Dunlop93} Dunlop, J. S., Taylor, G. L.,
Hughes, D. H., \& Robson, E. I. 1993, \mnras, 264, 455

\bibitem[Fukugita et al.(1995)]{Fukugita95} Fukugita, M., Shimasaku,
K., \& Ichikawa, T. 1995, \pasp, 107, 945

\bibitem[Gavazzi et al.(1998)]{Gavazzi98} Gavazzi, G., Catinella, B.,
Carrasco, L., Boselli, A., \& Contursi, A. 1998, \aj, 115, 1745

\bibitem[Hamabe \& Kormendy(1987)]{HK87} Hamabe, M., \& Kormendy,
J. 1987, in Structure and Dynamics of Elliptical Galaxies, ed. T. de
Zeeuw (Dordrecht, Holland: IAU), 379

\bibitem[Hooper et al.(1995)]{Hooper95} Hooper, E. J., Impey, C. D.,
Foltz, C. B., \& Hewett, P. C. 1995, \apj, 445, 62

\bibitem[Hooper et al.(1996)]{Hooper96} Hooper, E. J., Impey, C. D.,
Foltz, C. B., \& Hewett, P. C. 1996, \apj, 473, 746

\bibitem[Hooper et al.(1997)]{Hooper97} Hooper, E. J., Impey, C. D.,
\& Foltz, C. B. 1997, \apjl, 480, L95

\bibitem[Hutchings et al.(1984)]{Hutchings84} Hutchings, J. B.,
Crampton, D., \& Campbell, B. 1984, \apj, 280, 41

\bibitem[Hutchings \& Morris(1995)]{Hutchings95} Hutchings, J. B., \&
Morris, S. C. 1995, \aj, 109, 1541

\bibitem[Kellermann et al.(1989)]{Kellermann89} Kellermann, K. I.,
Sramek, R., Schmidt, M., Shaffer, D. B., \& Green, R. 1989, \aj, 98,
1195

\bibitem[Krist \& Burrows(1995)]{Krist95} Krist, J., \& Burrows,
C. J., 1995, Applied Optics, 34, 4951

\bibitem[Krist \& Hook(1999)]{Krist99} Krist, J., \& Hook, R., 1999,
The Tiny Tim User's Guide, (Baltimore: STScI), {\tt
http://www.stsci.edu/software/tinytim/tinytim.pdf}

\bibitem[Kristian(1973)]{Kristian73} Kristian, J. 1973, \apjl, 179,
L61

\bibitem[Kukula et al.(2001)]{Kukula01} Kukula, M. J., Dunlop, J. S.,
McLure, R. J., Miller, L., Percival, W. J., Baum, S. A., \& O'Dea,
C. P. 2001, preprint (astro-ph/0010007)

\bibitem[Lehnert et al.(1999)]{Lehnert99} Lehnert, M., van Breugel,
W., Heckman, T., \& Miley, G. 1999, \apjs, 124, 11

\bibitem[Magorrian et al.(1998)]{Magorrian98} Magorrian, J., et
al. 1998, \aj, 115, 2285

\bibitem[Malkan et al.(1984)]{Malkan84} Malkan, M. A., Margon, B., \&
Chanan, G. A. 1984, \apj, 280, 66

\bibitem[Metcalfe et al.(1998)]{Metcalfe98} Metcalfe, N., Ratcliffe,
A., Shanks, T., \& Fong, R. 1998, \mnras, 294, 147

\bibitem[McLeod \& Rieke(1994a)]{McLeod94a} McLeod, K. K., \& Rieke,
G. H. 1994a, \apj, 420, 58

\bibitem[McLeod \& Rieke(1994b)]{McLeod94b} McLeod, K. K., \& Rieke,
G. H. 1994b, \apj, 431, 137

\bibitem[McLure et al.(1999)]{McLure99} McLure, R. J., Kukula, M. J.,
Dunlop, J. S., Baum, S. A., O'Dea, C. P., \& Hughes, D. H. 1999,
\mnras, 308, 377

\bibitem[Nolan et al.(2001)]{Nolan01} Nolan, L. A., Dunlop, J. S.,
Kukula, M. J., Hughes, D. H., Boroson, T., \& Jimenez, R. 2001,
\mnras, 323, 417

\bibitem[Pence(1976)]{Pence76} Pence, W. 1976, \apj, 203, 39

\bibitem[Postman \& Lauer(1995)]{Postman95} Postman, M. \& Lauer,
T. R. 1995, \apj, 440, 28

\bibitem[Remy et al.(1997)]{Remy97} Remy, M., Surdej, J., Baggett, S.,
\& Wiggs, M. 1997, in 1997 HST Calibration Workshop, ed. S. Casertano
et al. (Baltimore: STScI), 374

\bibitem[Sabbey(1999)]{Sabbey99} Sabbey, C. N. 1999, PhD thesis, Yale
Univ. {\tt (ftp://www.astro.yale.edu/pub/sabbey/ thesis.ps.gz)}

\bibitem[Schaade et al.(2000)]{Schaade00} Schaade, D. J., Boyle,
B. J., \& Letawsky, M. 2000, \mnras, 315, 498

\bibitem[Schechter(1976)]{Schechter76} Schechter, P. 1976, \apj, 203,
297

\bibitem[Schlegel et al.(1998)]{Schlegel98} Schlegel, D. J., Finkbeiner, 
D. P., \& Davis, M. 1998, \apj, 500, 525

\bibitem[Smith et al.(1986)]{Smith86} Smith, E. P., Heckman, T. M.,
Bothun, G. D., Romanishin, W., \& Balick, B. 1986, \apj, 306, 64

\bibitem[Stockton \& MacKenty(1987)]{Stockton87} Stockton, A., \&
MacKenty, J. W. 1987, \apj, 316, 584

\bibitem[Surdej et al.(1997)]{Surdej97} Surdej, J., Baggett, S., Remy,
M., \& Wiggs, M. 1997, in 1997 HST Calibration Workshop,
ed. S. Casertano et al. (Baltimore: STScI), 386

\bibitem[V\'{e}ron-Cetty \& V\'{e}ron(1998)]{VCV} V\'{e}ron-Cetty,
M. P., \& V\'{e}ron, P. 1998, A Catalog of Quasars and Active Galactic
Nuclei (8th ed.; Garching: ESO)

\bibitem[Voit(1997)]{Voit97} Voit, M., ed. 1997, HST Data Handbook,
Vol. I (Version 3.0; Baltimore: STScI)

\bibitem[Yuan et al.(1998)]{Yuan98} Yuan, W., Brinkmann, W., Siebert,
J., Voges, W. 1998, \aap, 330, 108

\end{thebibliography}
\end {document}